\documentclass[journal]{IEEEtran}

\usepackage{cite}
\usepackage{amsmath,amssymb,amsfonts}
\usepackage{graphicx}
\usepackage{textcomp}
\usepackage{epsfig}
\usepackage{epstopdf}
\usepackage{color}
\usepackage{diagbox}
\usepackage{graphics}
\usepackage{stfloats}
\usepackage{multirow}
\usepackage{url}
\usepackage{subfigure}
\usepackage{algorithm}
\usepackage{algpseudocode}
\usepackage{amsmath}
\usepackage{caption}

\ifCLASSINFOpdf
\else
\fi
\begin{document}
%
\title{An Efficient Methodology to Identify Missing Tags in Large-Scale RFID Systems}
%
%

\author{Chu Chu,~\IEEEmembership{Graduate Student Member,~IEEE,}
        Rui Xu,
        Gang Li,
        Zhenbing Li,
       \\ and Guangjun Wen,~\IEEEmembership{Senior Member,~IEEE} 
      
\thanks{This work was supported in part by the National Natural Science Foundation of China under Project 61971113 and Project 61901095, in part by National Key R$\&$D Program under Project 2018YFB1802102 and Project 2018AAA0103203, in part by Guangdong Provincial Research and Development Plan in Key Areas under Project 2019B010141001 and Project 2019B010142001, in part by Sichuan Provincial Science and Technology Planning Program under Project 2018JY0246, Project 2019YFG0418, Project 2019YFG0120, and Project 2020YFG0039, in part by the Central Universities under Grant ZYGX2019Z022, 
in part by the Scholarship from China Scholarship Council (201906070019). (\emph{Corresponding author: Guangjun Wen})}
\thanks{Chu Chu, Rui Xu, Gang Li, Zhenbing Li, and Guangjun Wen are with the School of
Information and Communication Engineering, University of Electronic
Science and Technology of China, Chengdu 611731, China (email: chuchu$\_$824@163.com; xurui$\_$uestc@163.com; ligang1986718@163.com; thomaslizhenbing@163.com; wgj@uestc.edu.cn).}
}

%
%

\markboth{Journal of \LaTeX\ Class Files,~Vol.~14, No.~8, August~2015}%
{Shell \MakeLowercase{\textit{et al.}}: Bare Demo of IEEEtran.cls for IEEE Journals}
%



\maketitle

\begin{abstract}

Radio frequency identification (RFID) has been widely  has broad applications. One such application is
to use RFID to track inventory in warehouses and retail stores. In this application, timely identifying
the missing items is an ongoing engineering problem.
 A feasible solution to this problem is to map each tag to a time slot and verify the presence of a tag by comparing the status of the predicted time slot and the actual time slot. However, existing works are time inefficient because they only verify tags one by one in singleton slots but ignore the collision slots mapped by multiple tags. To accelerate the identification process, we use bit tracking to verify tags in collision slots and design two protocols accordingly. We first propose the Sequential String based Missing Tag Identification (SSMTI) protocol, which converts all time slots to collision slots and enables tags in each slot to reply to a designed string simultaneously. By using bit tracking to decode the combined string, the reader can verify multiple tags together. To improve the performance of SSMTI 
when most tags are missing, we further propose  the Interactive String based Missing Tag Identification (ISMTI) protocol. ISMTI improves the strategies of designing strings for each collided tag so that the reader can verify more tags using shorter strings than SSMTI.
Besides, ISMTI can dynamically adjust the verification mechanism according to the proportion of missing tags to maintain time efficiency. We also provide theoretical analysis for proposed protocols to minimize execution time and evaluate their performance through extensive simulations. Compared with state-of-the-art solutions, the proposed  SSMTI and ISMTI can reduce the time cost by as much as 39.74$\%$ and 68.87$\%$.


\end{abstract}

\begin{IEEEkeywords}
RFID, missing tag identification, bit tracking, time efficiency.
\end{IEEEkeywords}

%
\IEEEpeerreviewmaketitle

\section{Introduction}
\IEEEPARstart{R}{adio} frequency identification (RFID), as a compelling technology in the Internet of Things, is widely used in various applications, such as supply chain management \cite{supp1, supp2}, inventory control\cite{Chen2017Efficiently,chu}, and localization \cite{dingwei1, dingwei2}. In these applications, RFID can realize automated inventory management with its significant advantages of 
fast identification, high reliability, and long distance. Specifically,
an RFID system consists of at least one reader and multiple tags. Each tag in the system has a unique ID and is attached to a physical object. The RFID reader can collect information such as IDs from tags through wireless communication. Further, by tracking responded tags, the RFID reader can know the presence of corresponding objects.


Fast identification of missing objects is a critical task in RFID-enabled applications. According to statistics \cite{UK}, employee theft, vendor fraud, and management errors have become the main causes of inventory shrinkage in warehouses, retail stores, and supermarkets. In these scenarios, increasing delays in discovering and tracking the missing objects may cause substantial economic losses. Intuitively, to identify the missing objects, the RFID reader has to identify the missing tags. Therefore, this paper studies the challenging problem of missing tag identification, which can be 
formulated as follows: \textit{given a candidate tag set ${{S_{all}}}$, whose IDs are known by the reader, design a protocol that is capable of completely pinpointing the missing tags in ${{S_{all}}}$ using the minimum execution time, \textcolor{black}{which we define as time efficiency.} }

Although many approaches based on RFID have been proposed to identify the missing tags \cite{Su2017A,  Subit, Yubit, Liu2014Completely, PCMTI, CLS}, these approaches have the following limitations. 
First, the ID collection protocols are time-consuming \cite{EDFSA, Su2017A,  Subit}. In these protocols, each tag has to transmit its 96-bit ID that has been pre-stored in the reader. In addition, serious tag collisions cause many tags to transmit their IDs multiple times. 
Second, most existing missing tag identification protocols only verify tags one by one and therefore cannot meet the real-time requirements \cite{IIP, Liu2014Completely, PCMTI, CLS, MMTI,Suchao}.
 Specifically, existing works use the stored tags' ID and hash method to predict the slot selection of each tag in the frame and verify the presence of tags by comparing  the status of expected singleton slots with actual singleton slots. As shown in Fig. 1, the reader in existing works check whether slot 1 and slot 4 are empty or not during actual execution, so that only two tags ${t_1}$ and ${t_3}$ can be identified by the reader. 
 Although many efforts aim at increasing the proportion of singleton slots in the frame \cite{IIP, Liu2014Completely, MMTI},  they still waste many empty slots and collision slots. Moreover, most protocols ignore the impact of the missing rate, which is the proportion of missing tags to all candidate tags, on the execution time \cite {IIP, Liu2014Completely, PCMTI, MMTI,Suchao}. Therefore, the performance of such protocols deteriorates sharply at high missing rates. 

In this paper, we reduce the empty slots and extract the information hidden in the collision slots for tag verification. Specifically, we set a shorter frame length to map multiple tags in each slot. Then we separate the collided tags into different bits in a $w$-bit sequence, which we define as a string. These tags are expected to simultaneously transmit a string to the reader, and each tag sets one bit in its string to ``1'' and the other bits to ``0'' through Manchester encoding.

Accordingly, we first propose the Sequential String based Missing Tag Identification (SSMTI) protocol. SSMTI converts all slots to collision slots and  performs hash calculations to arrange each collided tag to a position of bit ``1'' in the string. After that, the reader enables tags to transmit their respective strings simultaneously in each collision slot. As shown in Fig. 1, tags ${t_1}$, ${t_2}$, and ${t_4}$ are respectively arranged to the first, second, and third bit in the string.
These three tags simultaneously transmit the strings, whose combination is expected to be ``XXX''. Here ``X'' is the collided bit. Since the reader only receives ``X0X'', it identifies that ${t_1}$ is missing and ${t_2}$, as well as ${t_4}$, are present.
Furthermore, we propose the Interactive String based Missing Tag Identification (ISMTI) protocol to accelerate the identification process under high missing rates. ISMTI arranges multiple tags to the same bit in the string so that the reader can identify more missing tags using a shorter string length. As shown in Fig. 1,
by checking the first bit of the string ``0X'', the reader identifies that ${t_1}$ and ${t_2}$ are missing.
We provide theoretical analysis for optimizing the parameters in our proposed protocols and conduct extensive simulation analysis to evaluate their performance under different working scenarios.  Numerical results, which match the analytical results well, demonstrate that the proposed protocols  can significantly reduce execution time by 39.74$\%$ and 68.87$\%$ when comparing with state-of-the- art solutions.

\begin{figure}
\centerline{\includegraphics[width=1\columnwidth]{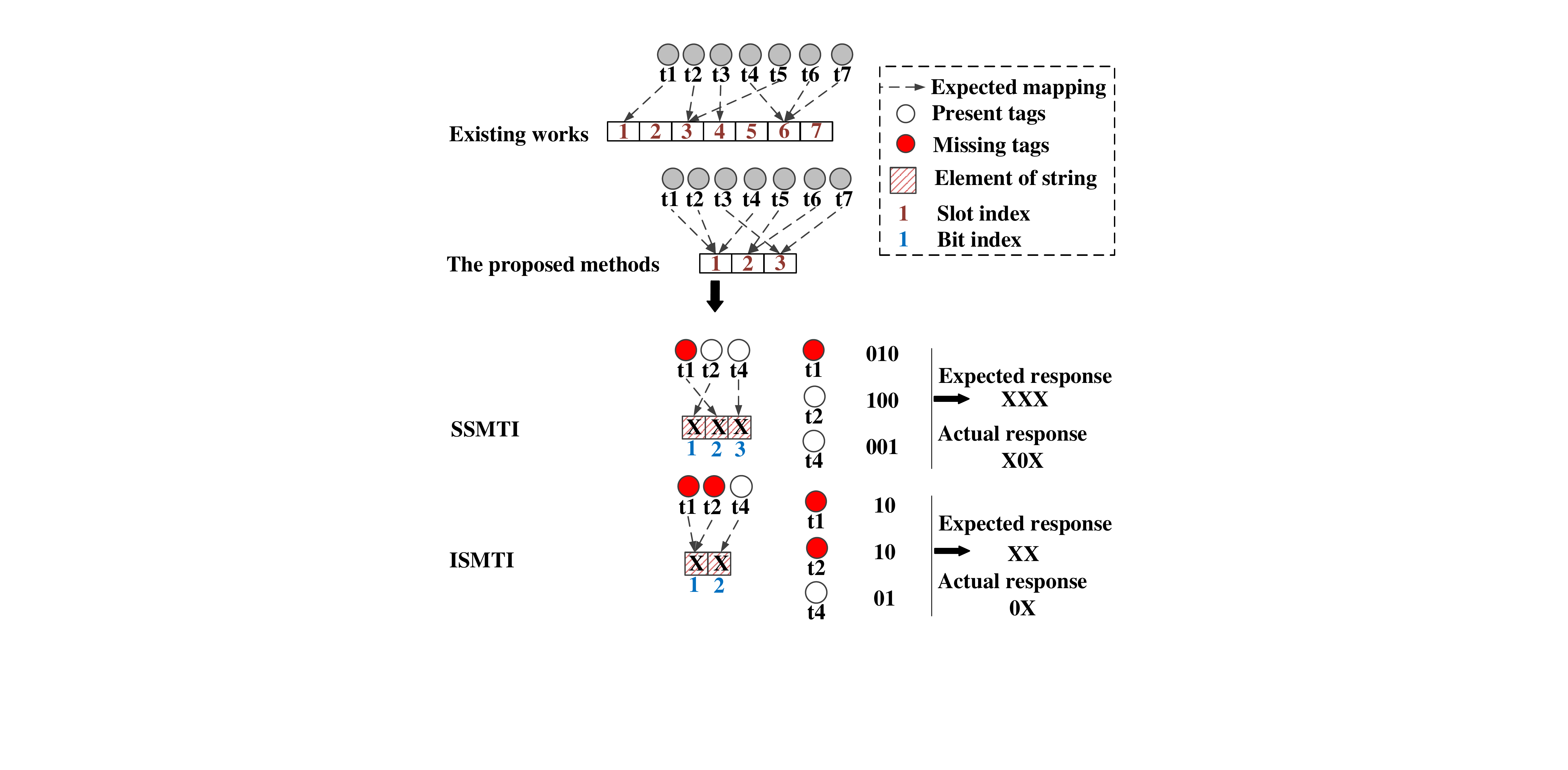}}
\caption{The proposed methods}
\label{fig1}
\end{figure}




The rest of the paper is organized as follows: Section II reviews the related works. Section III introduces the system models and formulates the main problem. The detailed process of our proposed protocols is illustrated in Section IV, Section V, and Section VI. Some discussions of the proposed protocols are described in Section VII. The simulation results and analyzes are presented in Section VIII. Section IX concludes the paper.
\section{Related Works}
Fast identifying the missing tags is critical to avoid substantial economic losses.
The solutions to missing tag identification can be
 classified into two categories: ID collection protocol \cite{Jian2016Idle, Subit, Yubit, EDFSA, DFSA2, Chen} and missing tag identification protocol \cite{Chen2017Efficiently, Rui2011Fast, IIP, MMTI, Liu2014Completely, PCMTI, CLS}.

Collecting the ID of all the tags under the reader's coverage is a direct way to identify the missing tags.  Existing ID collection protocols can be divided into two categories: Tree-based protocols \cite{Subit, Yubit} and Aloha-based protocols \cite{EDFSA, DFSA2, Chen}. In Tree-based protocols, the reader broadcasts a query command containing a string ``0'', and tags whose IDs start with ``0'' will respond. 
When the reader detects a collision, it appends a ``0'' or ``1'' to the previous string and queries the unidentified tags using this new string. When the reader successfully receives a reply, it identifies the involved tag. After the reader has identified all tags, the identification process in Tree-based protocols is terminated.
On the other hand, Aloha-based protocols divide the frame into multiple time slots and enable each tag to select a time slot randomly to reply to its ID. When two or more tags respond in the same time slot, the reader detects a collision and requires involved tags to reply again. When the reader successfully receives the tag ID in a time slot, it can identify the involved tag.  The identification process in Aloha-based protocols consists of multiple frames until all the tags are identified by the reader. In summary, ID collection protocols suffer from server tag collisions. Moreover, it is time-consuming for each tag to transmit the 96-bit ID that the reader already knows.

The missing tag identification protocols can be classified into two categories: probabilistic detection and deterministic identification. Probabilistic detection protocols \cite{Yu2019Finding, 7218577, Chen2017Probabilistic} aim at detecting whether or not any tags are missing under the reader's coverage with a predefined probability. However, these protocols are unable to completely identify the missing tags and therefore cannot solve the problem in this paper. On the other hand, deterministic identification protocols \cite{Chen2017Efficiently, Rui2011Fast, IIP, MMTI, Liu2014Completely, PCMTI, CLS} aim at determining the IDs of all the missing tags. \textcolor{black}{The basic idea of these protocols is to
pre-compute the expected status of each slot using the stored tag IDs and compare the status of the expected time slots with actual time slots. By checking the difference between the expected slots and actual slots, the reader can determine the involved missing tags.}

 The Iterative ID-free Protocol (IIP) \cite{IIP} was the early work to identify missing tags. In IIP, the reader checks the expected singleton slots to verify the presence of tags and increases the proportion of expected singleton slots using an additional hash function. 
 However, about half of the slots in the frame cannot be used by IIP. To improve the slot utilization of IIP, the Multi-hashing based Missing Tag Identification (MMTI) protocol \cite{MMTI} was proposed to turn most expected collision and empty slots into singleton slots for tag verification. However, to achieve this goal,
 the reader in MMTI has to spend much time broadcasting a long-bit indicator vector. 
 Similar to MMTI, the Slot Filter-based Missing Tag Identification (SFMTI) protocol \cite{Liu2014Completely} was proposed to turn a part of the expected two-collision slots and three-collision slots into singleton slots. Compared with IIP, SFMTI can achieve better time efficiency. 
Considering that the SFMTI protocol can only verify tags one by one, the Pair-Wise Collision-Resolving Missing Tag Identification (PCMTI) protocol \cite{PCMTI}, Coarse-grained inventory List based Stocktaking (CLS) protocol \cite{CLS}, and Collision Resolving based Missing Tag Identification (CR-MTI) protocol \cite{Suchao} were proposed. In PCMTI, the reader uses Manchester coding to verify two tags simultaneously in each expected two-collision slot that are mapped by two tags. On the other hand, the reader in CR-MTI verifies tags in expected singleton slots and part of the collision slots. However, many collision slots are still ignored by PCMTI and CR-MTI, thus limiting the performance of these protocols. Moreover, the above-mentioned identification protocols do not consider the impact of the 
missing rate on the execution time and become time-consuming as the missing rate increases.
The CLS protocol is designed to identify missing tags under high missing rates. Specifically, the reader checks each expected collision slot for tag verification. When an expected collision slot becomes empty during execution, the reader can identify multiple missing tags together. However, it is time-consuming for the reader to check all collision slots when most tags are present. 
In summary, existing works cannot effectively explore the expected collision slots for tag verification, which \textcolor{black}{has motivated} us to design new protocols.
\section{System Model and Problem Description}
\subsection{System Model}
A large-scale RFID system consists of a back-end server, an RFID reader, and $N$ RFID tags.
The back-end server has powerful capabilities of data processing and communicates with the reader via high-speed links. The RFID reader can communicate with tags within a certain range, which we define as the interrogation region. 
Each candidate tag has a unique 96-bit ID and is equipped with a hash function $H\left(  \cdot  \right)$. \textcolor{black}{Note that the hash based protocols assume that both the reader and tags share the common hash function, which is thoroughly studied \cite{IIP, Liu2014Completely, PCMTI, CLS}.} We use ${{{S_{all}}}} = \left\{ {{t_1},{t_2} \ldots {t_N}} \right\}$ to denote the set of candidate tags to be monitored in the RFID system. The set of tags that move out of the reader's interrogation region is called missing tags and is referred to as ${S_m}$. Accordingly, the remaining tag set is called present tags and is represented by ${S_p}$.  We use $N$, $M$, and $P$ to denote $|{S_{all}}|$, $|{S_{m}}|$, and $|{S_{p}}|$, where $|\cdot |$ stands for the cardinality of a set. By accessing the back-end server, the reader can generate an inventory list that records the IDs of all candidate tags. 


\subsection{Communication Model}
The communications between the reader and tags are assumed to be reliable and use the frame slotted Aloha protocol\cite{DFSA}.  \textcolor{black}{Specifically, the reader divides time into multiple frames, and each frame includes multiple time slots.   
At the beginning of a frame}, the reader initiates the communication by broadcasting the $Query$ command containing frame size $f$ and random seeds $r$. Upon receiving the reader's command, each tag calculates $H\left( {id,r} \right)\bmod {f}+1$ as the slot index to reply. 
According to tags' replies, the reader can divide the time slots into three types: empty slots with no tag reply, singleton slots with only one tag reply, and collision slots with multiple tag replies. Besides, the reader can also pre-compute the status of each slot using the stored IDs of $N$ tags. 
We refer to the time slots calculated by the reader as expected slots and the time slots during execution as actual slots.
The process of executing all the expected slots in a frame is called a round of identification, and the reader usually needs multiple rounds to identify all candidate tags.
\begin{figure*}
\centerline{\includegraphics[width=1.6\columnwidth]{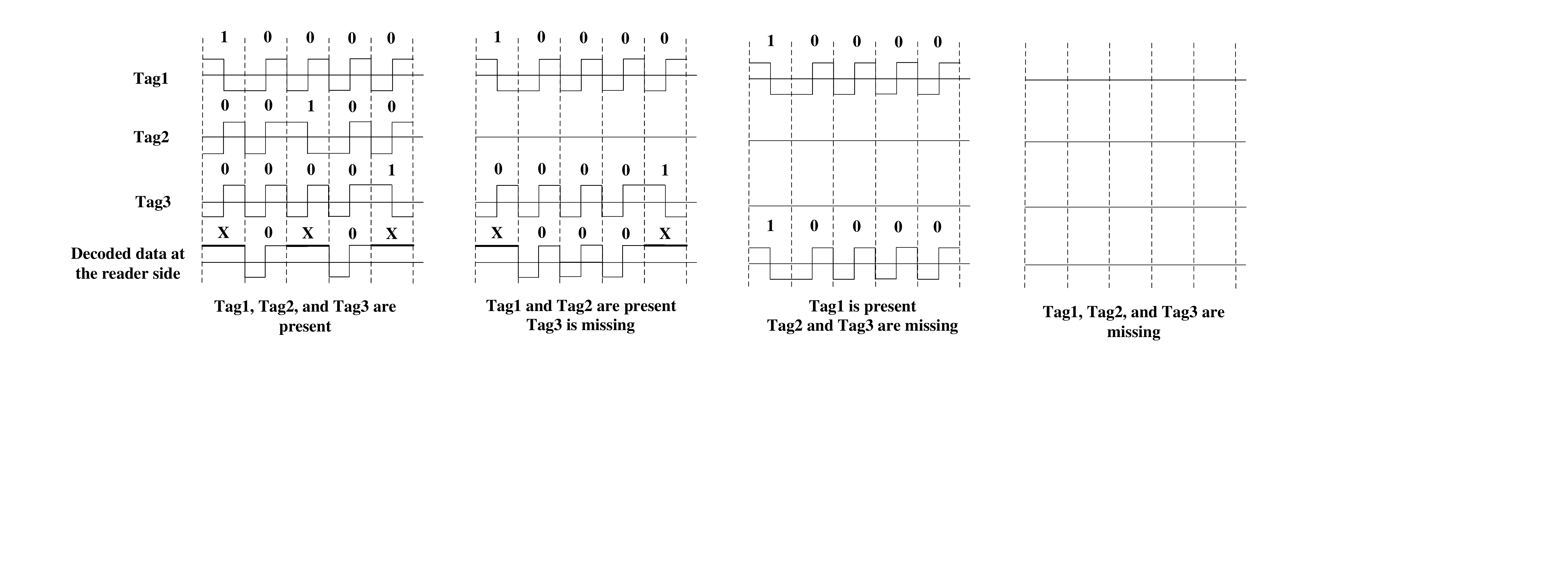}}
\caption{The use of Manchester encoding for missing tag identification}
\label{fig2}
\end{figure*}
\subsection{Problem Statement}

 Existing works only identify missing tags under a fixed missing rate in the system. However, the missing rate may vary in different application scenarios \cite{CLS}. For example, when we use a reader to perform high-frequency scans for the same region, candidate tags recorded in the inventory list can be updated periodically by the reader and therefore are usually close to the present tags. On the other hand, when we use the reader to scan the tags for the first time, the present tags might be much smaller than the candidate tags in the inventory list. 
Therefore, we concentrate on quickly identifying all the missing tags under various missing rates. Similar to previous works \cite{MMTI, Liu2014Completely, CLS}, the execution time is taken as the most critical performance metric. Generally, the problem can be summarized as follows: \textit{given the candidate set $S_{all}$ whose IDs are known by the reader, identify all missing tags $S_{m}$ in $S_{all}$ using the minimum execution time}.

\subsection{Bit Tracking}
Bit tracking is based on Manchester encoding, which defines a one-bit value as a voltage conversion within a fixed time. \textcolor{black}{In RFID systems, the bit tracking technology can detect the positions of collided bits in a string and therefore has been widely used in information collection protocols, anti-collision protocols, and unknown tag identification protocols \cite{infocollect1, Subit, Liu2014Efficient, libinbin2}. In these protocols, bit tracking requires tags to respond to the reader synchronously. Note that RFID systems usually operate at low rates and the transfer rate of a tag to the reader is 25 us \cite{MMTI,Suchao}. Since the synchronization offset for commercial RFID tags is no more than 1 us \cite{bit2}, a bit duration offset of $4\%$ does not have much negative on bit tracking \cite{tongbu} and
the clocks of the tags can be well synchronized via the signal received from the reader \cite{infocollect1, Subit}. 
Moreover, the practical experiment of bit synchronization with USRP and WISP also verified that the bit tracking in the UHF RFID system is feasible \cite{tongbu}.
Therefore, we assert that the signals from each tag can be well synchronized on bit-level \cite {infocollect1, Subit, Liu2014Efficient, libinbin2} and introduce bit tracking technology in this paper to accelerate tag identification and the proposed method.}

Different from most missing tag identification protocols that only verify each tag in the expected singleton slot, we use bit tracking to explore the information hidden in  collision slots for tag verification.
\textcolor{black}{As shown in Fig. 2, we assume that an expected collision slot is mapped by Tag 1, Tag 2, and Tag 3.  The reader separates these tags into different bits in a five-bit string, which is represented by $W$.  Since Tag 1 is arranged to the first bit, this tag sets the value of this bit to ``1'' and the value of the other bits to ``0''.
Similarly, Tag 2 and Tag 3 are arranged to the third and fifth bit. These three tags are expected to transmit their strings to the reader simultaneously. By decoding the first, third, and fifth bits of the combined string, the reader can determine whether or not any tags are missing. Specifically, When the reader detects a collided bit or a bit value of ``1'', the tag arranged to this bit can be identified as the present tag. On the other hand, when the reader detects a bit value of ``0'' or receives no signal, it can identify the tag arranged to this bit as 
a missing tag. Following this idea, we design three protocols to identify the missing tags quickly.}

\section{Sequential String based Missing Tag Identification Protocol}
In the most related CR-MTI protocol \cite{Suchao}, multiple tags can be identified by the reader in the expected collision slots only when each tag obtains a unique value in the string. However, due to the probabilistic nature of the hash function, many expected collision slots cannot be resolved.
Moreover, to resolve the collision slots, the string length $w$ has to be larger than the involved tags, resulting in multiple bits in the string are not arranged by any tags. 
To address this problem,  we propose the Sequential  String-based Missing Tag Identification (SSMTI) protocol. SSMTI converts all the expected slots into resolvable collision slots and uses short indicator vectors to make each bit in the string corresponding to a tag.
SSMTI consists of four stages:
main filter vector construction stage, collided tags reconciling stage, candidate tags sequencing stage, and present tag verification stage.
The reader executes the first three stages round by round until each candidate tag has been arranged a unique value. After that, the reader begins the fourth stage and checks the response of each slot.
In what follows, we will describe the SSMTI protocol.

\subsection{Protocol Description of SSMTI}
\subsubsection{Main Vector Construction Stage}

In this stage, the reader maps each tag to a bit and constructs a main vector accordingly.
Specifically, the reader constructs the main vector ${F_2}$ with ${f_2}$ elements as follows: if a bit is selected by one tag, the reader sets the corresponding elements in ${F_2}$ to ``1''; 
if a bit is selected by two tags, the reader sets the corresponding elements in ${F_2}$ to ``2''; otherwise, the reader sets the corresponding elements in ${F_2}$ to ``0''.  An example is illustrated in Fig. 5, where the reader constructs the main vector ${F_2}=200102$.  

\subsubsection{Collided Tags Reconciling Stage}
For each bit that is selected by two tags, the reader generates another seed $r2$ and calculates $H\left( {id,r2} \right)\bmod 2$ using the involved tags' IDs. If each tag obtains a unique value of $0$ or $1$, we say this bit is reconcilable. After that, the reader updates the value of the corresponding elements in ${F_2}$: if a bit is reconcilable, the reader updates the corresponding element in ${F_2}$ to ``4''; if a bit is unreconcilable, the reader updates the corresponding element in ${F_2}$ to ``3''.  An example of the reconciling stage is shown in Fig. 5. Since $H(id_{t_4},{{r_2}})\bmod 2=0$ and  $H(id_{t_8},{{r_2}})\bmod 2=1$,  
the first bit is reconcilable. On the other hand, ${t_7}$ and ${t_9}$ obtain the same value of $0$. Therefore, the sixth bit is unreconcilable.

After that, the reader constructs an append filter vector $A$ with $a$ elements based on the following rules:

(i) If ${F_2}\left[ i \right]{\text{ = }}1$, which means that the $i$th bit is selected by one tag, \textcolor{black}{the reader will set the ${a _i}+1 $th element of vector $A$ to ``0''}, 
where ${a _i}$ is the number of elements ``1'' and ``4'' proceeding the ${F_2[i]}$.

(ii) If $F_2\left[ i \right]{\text{ = }}4$, which means that the $i$th bit is selected by two tags and is reconcilable, the reader will set the ${a _i}+1$th element of vector $A$ to ``1''.

As illustrated in Fig. 5, since ${F_2}\left[ 1 \right]{\text{ = }}4$ and ${a _1}=0$, the reader sets $A[1]=1$. On the other hand, since ${F_2}\left[ 4 \right]{\text{ = }}1$ and ${a _4}=1$, the reader sets $A[2]=0$. 


\subsubsection{Candidate Tags Sequencing Stage}
\textcolor{black}{In this stage,
the reader combines the main vector ${F_2}$ and the append vector $A$ to construct the indicator vector ${{V_2}}$.}
Specifically, if ${F_2}[i]=1$ and $A[{a _i}+1]=1$ , the reader will set the $i$th element of ${V_2}$ to ``11''; if ${F_2}[i]=1$ and $A[{a _i}+1]=0$, the reader will set the $i$th element of ${V_2}$ to ``10''; otherwise, the reader sets the $i$th element of ${V_2}$ to ``0''. As shown in Fig. 5, since ${F_2}[1]=1$ and $A[0+1]=1$, the reader sets ${V_2}[1]=11$. Similarly, the reader sets ${V_2}[4]=10$ 
because ${F_2}[4]=1$ and $A[1+1]=0$. 
In our construction method, only the elements corresponding to reconcile 2-collision slot and singleton slot are encoded by 2 bits in ${V_2}$. 
In the simulation, we will evaluate that this constructed method can reduce the number of bits transmitted by the reader. 

After constructing the indicator vector, the reader broadcasts ${V_2}$ and parameters, including ${f_2}$, $a$, $r_1$, $r_2$,  ${w}$, and $\mu $ to all tags. Here $\mu $ is the number of tags that have received a unique number before the current execution round.

Upon receiving the indicator vector and parameters, each present tag, say ${t_c}$, obtains an index $i$ by calculating $H\left( {i{d_{{t_c}}},{r_1}} \right)\bmod {f_2}+1$. Then ${t_c}$ checks the indicator vector ${V_2}$.  Let ${\chi_{10}}$
and ${\chi_{11}}$ be the number of ``10'' and ``11'' before the $i$th element in ${{V_2}}$. ${t_c}$ will store ${\chi_{10}}$
and ${\chi_{11}}$, and determine where and how to reply to the reader based on the following rules:

 (iii) If ${{V_2}}\left[ i \right]{\text{ = }}0$, which means that the $i$th bit is unreconcilable, ${t_c}$ will wait for the indicator vector in the next round.

(iv) If ${{V_2}}\left[ i \right]{\text{ = }}10$, which means that the $i$th bit is selected by one tag,  ${t_c}$ will calculate $\chi  
=\mu + {\chi _{10}}+ 2 \times {\chi _{11}} + 1$ as its unique value and wait for the beginning of the fourth stage. 

(v) If ${{V_2}}\left[ i \right]{\text{ = }}11$, which means that the $i$th bit is reconcilable, ${t_c}$ will perform $H\left( {id,r2} \right)\bmod 2$. If the hashing result is ``0'', ${t_c}$ will calculate $\chi  = \mu + {\chi _{10}}+2 \times {\chi _{11}} + 1$ as its unique value. If the hashing result is ``1'', ${t_c}$ will calculate $\chi  = {\mu}+{\chi _{10}} + 2 \times {\chi _{11}}+ 2$ as its unique value. After that, ${t_c}$ will wait for the beginning of the fourth stage.  

As illustrated in Fig. 5, since ${V_2}(2)={V_2}(3)={V_2}(5)={V_2}(6)=0$, ${t_1}$, ${t_2}$, ${t_3}$, ${t_6}$, ${t_7}$, ${t_9}$, ${t_{10}}$, and ${t_{11}}$ wait for the indicator vector in the next round. On the other hand,
since ${V_2}(1)=11$, ${t_4}$ and ${t_8}$ perform $H(id,{{r_2}})\bmod 2$. Then ${t_4}$ calculates ${\chi}=0+0+2{\times }0+1$ and ${t_8}$ calculates ${\chi}=0+0+2{\times }0+2$. Note that in the example of Fig. 5, we assume no tag has been arranged a unique number before the current round. Therefore, $\mu =0$.
Since ${V_2}(4)=10$ and the number of ${\chi_{11}}$ before the fourth element of ${V_2}$ is 1, ${t_5}$ calculates ${\chi}=0+0+2{\times }1+1$. Both ${t_4}$, ${t_5}$, and ${t_8}$ keep silent and wait for the beginning of the fourth stage.

\subsubsection{Present Tag Verification Stage}
After each tag has obtained a unique value $\chi$, the reader begins the fourth stage. In this stage, each present tag ${t_c}$ will calculate $ \left\lceil {\frac{\chi }{w}} \right\rceil $ as its actual slot to reply. Meanwhile, the tag will calculate $j= \chi \bmod w$  to sets the $j$th bit to ``1'' and all other $w-1$ bits to ``0'' in its string. In Fig. 5, ${t_5}$ calculates $ i= \left\lceil {\frac{3 }{6}} \right\rceil =1$ and $j = 3 \bmod 6 =3$. Therefore, ${t_5}$ transmits a 6-bit string ``001000'' in slot 1.

After replying, ${t_c}$ will keep silent and not participate in the subsequent identification process.

\textcolor{black}{On the reader side, the reader checks each receiving string, whose bits are all expected to ``1''. If the reader detects a  difference, the corresponding tags are identified as missing. As illustrated in Fig. 5, since the reader detects ``XXX0X0'' in slot 1 and ``X0X000'' in slot 2, the reader identifies that ${t_1}$,  ${t_3}$, ${t_2}$, ${t_7}$, and ${t_9}$ are missing. Meanwhile, the other tags are identified as present.}

\begin{figure}
\centerline{\includegraphics[width=1\columnwidth]{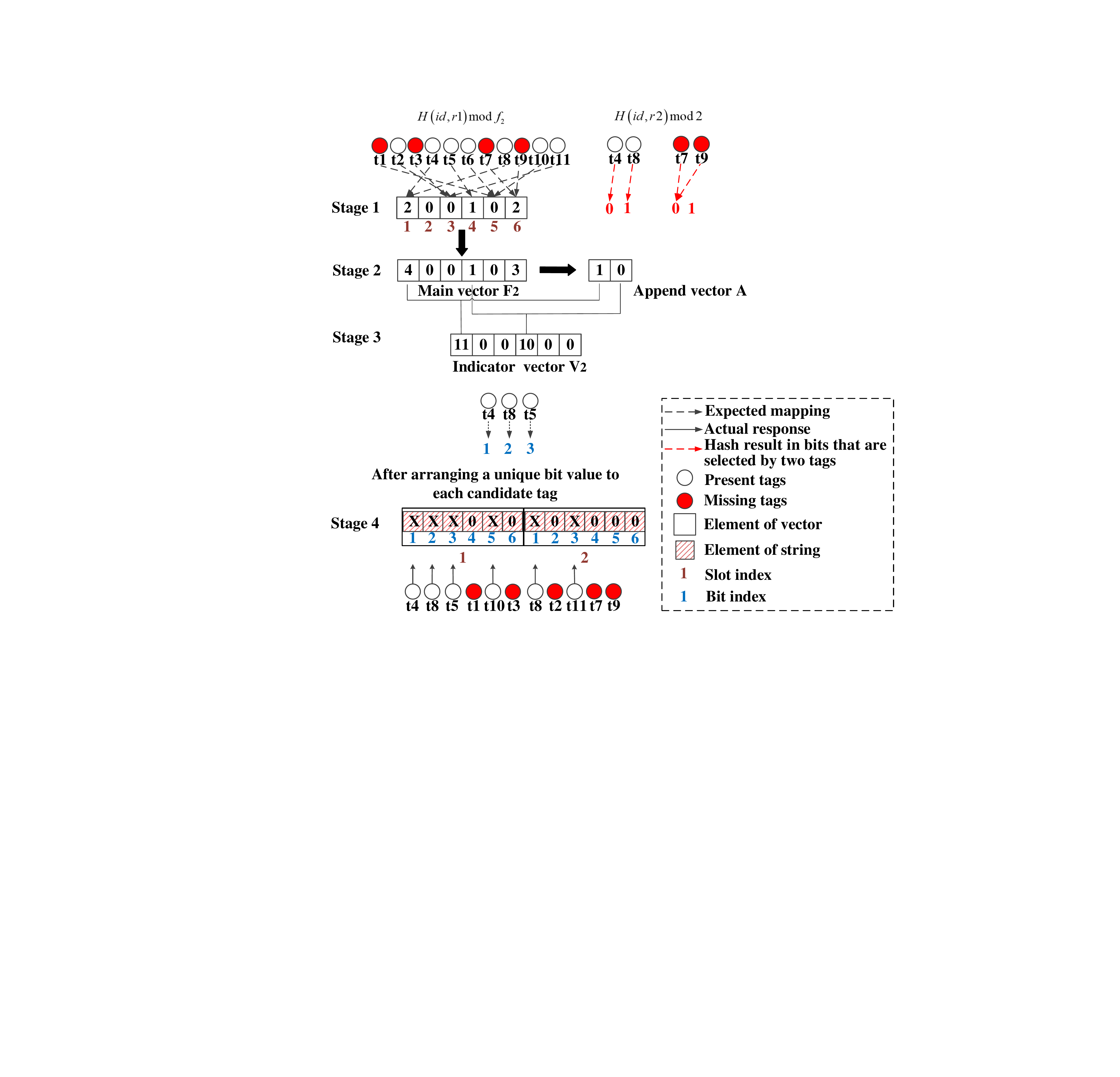}}
\caption{Flow diagram of SSMTI protocol}
\label{fig5}
\end{figure}
\subsection{Parameter Optimization}
We first determine the optimal main vector length ${f_2}$ for the first three stages.  The tag that has obtained a unique value is called the arranged tag.
Let ${P_{SS}}$ denote the probability that a candidate tag can be successfully arranged to a unique value. Then we have
\begin{equation}
{P_{SS}} = P_{SS}^1 + P_{SS}^2
\end{equation}
where ${P_{SS}^1}$  is the probability that a bit is mapped by only one tag and ${P_{SS}^2}$ is the probability that a bit is mapped by two tags and is reconcilable. Here we calculate ${P_{SS}^1}$  as
\begin{align}
 P_{SS}^1 &= \left( {\begin{array}{*{20}{c}}
   {{N^ * }}  \\
   1  \\
\end{array}} \right)\left( {\frac{1}{{f_2}}} \right){\left( {1 - \frac{1}{{f_2}}} \right)^{{N^ * } - 1}} \nonumber\\
  &\approx \frac{{{N^ * }}}{{f_2}}{e^{ - \frac{{{N^ * } - 1}}{{f_2}}}}.
 \end{align}

For a bit that is mapped by two tags, the probability that it can be reconciled is $\frac{{2!}}{{{2^2}}}$. Hence, we calculate ${P_{SS}^2}$ as
\begin{align}
 P_{SS}^2 &= \left( {\begin{array}{*{20}{c}}
   {{N^ * }}  \\
   2  \\
\end{array}} \right){\left( {\frac{1}{{f_2}}} \right)^2}{\left( {1 - \frac{1}{{f_2}}} \right)^{{N^ * } - 2}} \times \frac{{2!}}{{{2^2}}} \nonumber\\
  &\approx \frac{{{N^ * }\left( {{N^ * } - 1} \right)}}{{4{{f_2}^2}}}{e^{ - \frac{{{N^ * } - 2}}{{f_2}}}} \nonumber.
\end{align}

Let $\phi_{SS} $ denote the expected number of bits that can be used for unique value arrangement in each round of the first three stages, and let ${\phi_{SS}^k}$ denote the number of bits selected by $k$ tags. We can calculate $\phi_{SS} $ as
\begin{align}
 \phi_{SS}  &= {\phi _{SS}^1} + {\phi _{SS}^2} \nonumber \\
 & = {f_2} \times P_{SS}^1 + {f_2} \times P_{SS}^2 \nonumber\\
  &= {f_2}p_2{e^{ - p_2}} + \frac{1}{4}{f_2}{p_2^2}{e^{ - p_2}}
\end{align}
 where $\frac{{{N^ * }}}{{f_2}}=p_2$. Let $\aleph_{SS}$ denote the number of arranged tags in each round of the first three stages and ${\aleph_{SS}^k}$ denote the number of arranged tags in the bits that are selected by $k$ tags. We can calculate $\aleph_{SS}$ as
\begin{align}
 \aleph_{SS}  &= {\aleph _{SS}^1} + {\aleph _{SS}^2} \nonumber \\
  &= {f_2} \times P_{SS}^1 + {f_2} \times 2 \times P_{SS}^2 \nonumber \\
  &= {f_2}p_2{e^{ - p_2}} + \frac{1}{2}{f_2}{p_2^2}{e^{ - p_2}}.
\end{align}

Let ${T_{SS1}}$ denote the execution time of each round in the first three stages. Note that the length of the indicator vector ${V_2}$ is ${f_2} +a$. Hence, we can calculate ${T_{SS1}}$ as 
\begin{align}
 {T_{SS1}} &= \left[ {\frac{{({f_2} + a)}}{{96}}} \right] \times {t_{tag}} \nonumber\\
  &= \left[ {\frac{{({f_2} + {\phi _{1,SS}} + {\phi _{2,SS}} )}}{{96}}} \right] \times {t_{tag}}.
\end{align}

\textcolor{black} {Combing (13),  (14), and (15), we can calculate the efficiency to arrange a unique value to each tag in the first three stages as}
\begin{align}
 \frac{\aleph_{SS} }{T_{SS1}} &= \frac{{{\aleph _{SS}^1} + {\aleph _{SS}^2}}}{{\left[ {\frac{{({f_2} + {\phi _{1,SS}} + {\phi _{2,SS}} + 1)}}{{96}}} \right] \times {t_{tag}}}} \nonumber\\
  &\approx \frac{{p_2{e^{ - p_2}} + \frac{1}{2}{p_2^2}{e^{ - p_2}}}}{{\left( {p_2{e^{ - p_2}} + \frac{1}{4}{p_2^2}{e^{ - p_2}} + 1} \right) \times \frac{{{t_{tag}}}}{{96}}}}.
\end{align}

The objective of the first three stages is to maximize the efficiency. Hence we calculate the derivative of the efficiency with respect to $p_2$ as
\begin{align}
{\left( {\frac{\aleph_{SS} }{T_{SS1}}} \right)^\prime } = \frac{{\left\{ \begin{array}{l}
 \left( {1 - \frac{1}{2}{p_2^2}} \right) \times \left[ {\left( {\frac{1}{4}{p_2^2} + p_2} \right){e^{ - p_2}} + 1} \right] \nonumber\\
  - \left( {\frac{1}{2}{p_2^2} + p_2} \right) \times \left( {1 - \frac{1}{2}p_2 - \frac{1}{4}{p_2^2}} \right) \\
 \end{array} \right\}{e^{ - p_2}}}}{{{{\left( {p_2{e^{ - p_2}} + \frac{1}{4}{p_2^2}{e^{ - p_2}} + 1} \right)}^2} \times \frac{{{t_{tag}}}}{{96}}}}.
\end{align}

 We can obtain an optimal ${p_{opt}}$ to satisfy ${\left( {\frac{\aleph_{SS} }{T_{SS1}}} \right)^\prime }{\rm{ = }}0$. When ${p_{opt}} \approx 1.50$, ${{\aleph_{SS} }/{T_{SS1}}}$ is maximized. Therefore we obtain the optimal size of the main vector ${f_{opt}} = {{{N^*}} \mathord{\left/
 {\vphantom {{{N^ * }} {1.5}}} \right.
 \kern-\nulldelimiterspace} {1.5}}$ and the execution time of the first three stages is
${{NT} \mathord{\left/
 {\vphantom {{NT} \aleph }} \right.
 \kern-\nulldelimiterspace} \aleph } = {N \mathord{\left/
 {\vphantom {N {16.0448}}} \right.
 \kern-\nulldelimiterspace} {16.0448}} \approx 0.0623N$, where $N$ is the number of total candidate tags. 



 Let ${T_{SS2}}$ denote the execution time in the present tag verification stage. ${T_{SS2}}$ includes the time for all tags to reply to the $w$-bit strings. Since each tag is arranged to a unique value, the total number of strings is equal to $\left\lceil {\frac{N}{w}} \right\rceil $. Recall that the time to transmit one string is ${t_w} = 0.4 + (w - 1) \times 0.025$, we can calculate  ${T_{SS2}}$  as 
  \begin{align}
 {T_{SS2}} &= \left\lceil {\frac{N}{w}} \right\rceil  \times {t_w} \nonumber \\
  &= \left\lceil {\frac{N}{w}} \right\rceil  \times \left( {0.4 + (w - 1) \times 0.025} \right)
\end{align}
where $\left\lceil {} \right\rceil $ is the ceiling function.
Therefore, we can calculate the efficiency of identifying each tag in the fourth stage as 
  \begin{align}
\frac{N}{{{T_{SS2}}}} \approx \frac{w}{{\left( {0.4 + (w - 1) \times 0.025 } \right)}}.
  \end{align}

Since $w > 0$, ${N}/{{{T_{SS2}}}}$  is a monotonically increasing function with respect to $w$. 
Intuitively, the larger the $w$ value, the more tags can be identified in each string. In this paper, we set $w=96$ as the length of each string in SSMTI because the maximum length of the vector should not be greater than 96 bits \cite{IIP, MMTI, Liu2014Efficient, tongbu}. Therefore, the execution time of SSMTI is approximately $(0.0623N+\left\lceil {\frac{N}{96}} \right\rceil  \times 2.375)$ ms.


\section{Interactive String based Missing Tag Identification Protocol}
In the proposed SSMFI protocol, each bit in the string corresponds to a tag to be verified. However, when the missing rate is high, we can arrange multiple tags to the same bit value in a string to further improve the time efficiency of SSMFI. Following this idea, we propose the Interactive String-based Missing Tag Identification (ISMFI) protocol, which consists of two stages: the present tag response stage and the missing tag verification stage.
ISMTI executes two stages round by round until all the tags are identified. The detailed operations of these two phases are presented below.

\subsection{Present Tag Response Stage}
In this stage, the reader broadcasts the parameters ${f_3}$, $r$, and $w$ to all the tags.
 Subsequently, the reader predicts the slot and bit mapping of each tag using tags' IDs. Accordingly, the reader constructs an expected  vector ${EV}$ consisting of ${f_3}$ elements. 
 Specifically, the element's value of ${EV}$ is encoded by ``0'' if the corresponding bit is not mapped by any tags in the string; ``1''  if the corresponding bit is mapped by one tag in the string; ``2''  if the corresponding bit is mapped by multiple tags in the string. As depicted in Fig. 6, the reader constructs the expected frame vector ${EV}$=022222.

Once receiving the parameters, each tag ${t_c}$ will calculate $i = \left\lceil {{{\left[ {H\left( {i{d_{{t_c}}},r} \right)\bmod f+1} \right]} \mathord{\left/
 {\vphantom {{\left[ {H\left( {i{d_{{t_c}}},r} \right)\bmod f} \right]} w}} \right.
 \kern-\nulldelimiterspace} w}} \right\rceil $ as the actual slot to reply.  Meanwhile, ${t_c}$ will calculate $j=\left[ {H\left( {i{d_{tc}},r} \right)\bmod f} \right]\bmod w +1$ by setting the $j$th bit to ``1'' and all other $w-1$ bits to ``0''. Subsequently, ${t_c}$ will transmit its string in the $i$th slot to the reader. For example, since ${t_8}$ calculates $\left\lceil {{{\left[ {H\left( {i{d_{{t_8}}},r} \right)\bmod 6+1} \right]} \mathord{\left/
 {\vphantom {{\left[ {H\left( {i{d_{{t_6}}},r} \right)\bmod 6} \right]} 3}} \right.
 \kern-\nulldelimiterspace} 3}} \right\rceil = 1 $ and $\left[ {H\left( {i{d_{t_8}},r} \right)\bmod 6} \right]\bmod 3 +1 =3$, ${t_8}$ selects slot 1 to transmit its string $001$.

\subsection{Missing Tag Verification Stage}
After receiving the strings from all the present tags, the reader can generate an actual vector ${AV}$ consisting of ${f_3}$ elements. \textcolor{black} {Specifically, for each bit of strings received by the reader, if the reader detects ``1'' or ``X'', the reader will set the corresponding element in ${AV}$ to ``1''; Otherwise, the reader will set the corresponding element in ${AV}$  ``0'.} As some tags are missing, some elements in ${EV} $ with a value of ``1'' or ``2'' will become ``0'' in ${AV}$. Therefore, the reader can identify the missing tags by comparing ${EV}$ and ${AV}$. As shown in Fig. 6, since the reader detects ``0'' or no signal in the fourth, fifth, 
sixth bit of ${AV}$, the reader identifies that ${t_3}$, ${t_5}$, ${t_7}$, ${t_4}$, ${t_9}$, ${t_{10}}$, and ${t_{11}}$ are missing.

Then, the reader constructs the indicator vector ${V_3} $ according to the following rules:

 (i) If ${EV}\left[ i \right] = 0$, ${V_3}\left[ i \right] = 0$.

(ii) If ${EV}\left[ i \right] = 1$ or $2$, and ${AV}\left[ i \right] = 0$, ${V_3}\left[ i \right] = 1$.

(iii) If ${EV}\left[ i \right] = 1$  and ${AV}\left[ i \right] = 1$, ${V_3}\left[ i \right] = 1$.

(iv) If ${EV}\left[ i \right] = 2$  and ${AV}\left[ i \right] = 1$, ${V_3}\left[ i \right] = 0$.

After that, the reader broadcasts ${V_3}$ to all the tags. Each tag ${t_c}$ checks the value of its corresponding bits in ${V_3}$. If the value happens to be ``1'', ${t_c}$ will keep silent and not participate in the subsequent identification process; otherwise, ${t_c}$ will wait for the beginning of the next round. As shown in Fig .6, ${t_1}$, ${t_2}$, ${t_6}$, and ${t_8}$ are expected to participate in the next round.   


\begin{figure}
\centerline{\includegraphics[width=1\columnwidth]{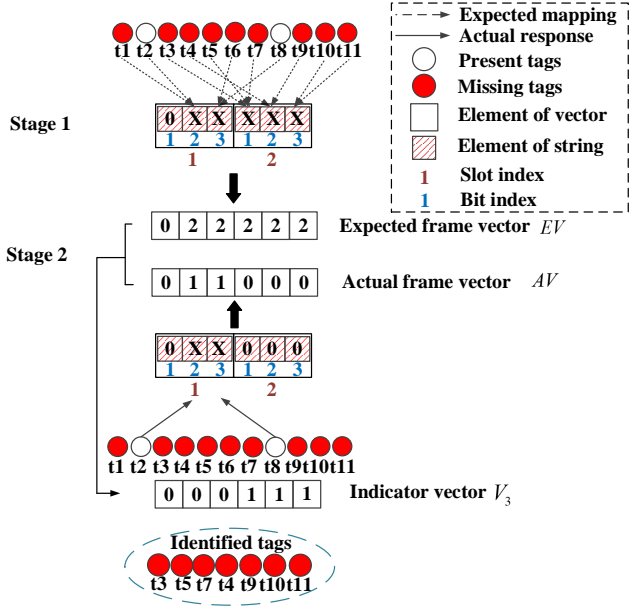}}
\caption{Flow diagram of ISMTI protocol}
\label{fig6}
\end{figure}
\subsection{Parameter Optimization}
The execution time of the proposed ISMTI protocol is influenced by the length of the vector ${f_3}$ and string $w$. In this subsection,  we determine the optimal value of ${f_3}$ and string length $w$ to achieve the best time efficiency.

In ISMTI, a tag can be successfully identified only when the bit in the string is arranged to one tag or multiple missing tags. This probability is denoted by ${P_{IS}}$ and can be written as
 \begin{align}
{P_{IS}} = P_{IS}^1 + P_{IS}^k
 \end{align}
\textcolor{black} {where ${P_{IS}^1}$ is the probability that a bit is mapped by one tag and this probability is the same as (12); ${P_{IS}^k}$ is the probability a the bit is mapped by $k$ missing tags.} Let $q$ denote the current missing rate in each round, the probability that $k$ tags are missing is equal to ${q^k}$. Hence, we calculate ${P_s^k}$ as 
 \begin{align}
P_{IS}^k = \sum\limits_{k = 2}^{{N^ * }} {\left( {\begin{array}{*{20}{c}}
   {{N^ * }}  \\
   k  \\
\end{array}} \right){{\left( {\frac{1}{{f_3}}} \right)}^k}{{\left( {1 - \frac{1}{{f_3}}} \right)}^{{N^ * } - k}} \times {q^k}}.
 \end{align}

Let $\aleph_{IS}$ denote the total number of tags that can be successfully identified in each round and ${\aleph _{IS}^k}$ to represent the number of tags that can be identified in the bit that are mapped by one tag or $k$ missing tags. We can calculate ${\aleph _{IS}^k}$ $(k>1)$ as
 \begin{align}
 \sum\limits_{k = 2}^{{N^ * }} {{\aleph _{IS}^k}}  &= {f_3}\sum\limits_{k = 2}^{{N^*}} {k\left( {\begin{array}{*{20}{c}}
   {{N^*}}  \\
   k  \\
\end{array}} \right){{\left( {\frac{1}{{f_3}}} \right)}^k}{{\left( {1 - \frac{1}{{f_3}}} \right)}^{{N^*} - k}} \times {q^k}}  \nonumber\\
  &= {f_3}\left[ {\sum\limits_{k = 1}^{{N^*}} {k\left( {\begin{array}{*{20}{c}}
   {{N^*}}  \\
   k  \\
\end{array}} \right){{\left( {\frac{1}{{f_3}}} \right)}^k}{{\left( {1 - \frac{1}{{f_3}}} \right)}^{{N^*} - k}}{q^k}} } \right] \nonumber\\
  &- {f_3}P_{IS}^1q \nonumber\\
  &= {f_3}\left[ {\sum\limits_{k = 1}^{{N^*}} {\left( {\begin{array}{*{20}{c}}
   {{N^*} - 1}  \\
   {k - 1}  \\
\end{array}} \right){{\left( {\frac{q}{{f_3}}} \right)}^{k - 1}}{{\left( {1 - \frac{1}{{f_3}}} \right)}^{{N^*} - k}}} } \right] \nonumber\\
  &\times \frac{{{N^*}q}}{{f_3}} - {f_3}P_{IS}^1q.
 \end{align}

 According to the binomial theorem,  we rewrite (21) as 
 \begin{align}
 \sum\limits_{k = 2}^{{N^*}} {{\aleph _{IS}^k}}  &= {f_3}\left[ {\frac{{{N^*}q}}{{f_3}}{{\left( {\frac{q}{{f_3}} + 1 - \frac{1}{{f_3}}} \right)}^{{N^*} - 1}} - P_{IS}^1q} \right] \nonumber\\
  &= {f_3}\left[ {\frac{{{N^*}q}}{{f_3}}{{\left( {1 - \frac{{1 - q}}{{f_3}}} \right)}^{{N^*} - 1}} - P_{IS}^1q} \right] \nonumber\\
  &\approx {f_3}\left( {{{p_3q}}{}{e^{ - p_3\left( {1 - q} \right)}} - p_3q{e^{ - p_3}}} \right)
 \end{align}
where  $p_3=\frac{{{N^ * }}}{{f_3}}$. Hence we can obtain $\aleph_{IS}$ as
 \begin{align}
 \aleph_{IS}  &= {\aleph _{IS}^1} + \sum\limits_{k = 2}^{{N^ * }} {{\aleph _{IS}^k}} \nonumber \\
  &= {f_3} \times \left( {p_3{e^{ - p_3}} + {{p_3q}}{}{e^{ - p_3\left( {1 - q} \right)}} - p_3q{e^{ - p_3}}} \right).
 \end{align}

Let ${T_{IS}}$ denote the execution time in each round of ISMTI. ${T_{IS}}$ includes the transmission time of the ${f_3}$-bit string and the ${f_3}$-bit indicator vector ${V_3}$. Hence, we calculate ${T_{IS}}$ as 
 \begin{align}
{T_{IS}} = \left\lceil {\frac{{f_3}}{{96}}} \right\rceil  \times {t_{w}} + \left\lceil {\frac{{f_3}}{{96}}} \right\rceil  \times {t_{tag}}.
 \end{align}

 Combing  (23) and  (24), we can obtain the efficiency as 
 \begin{align}
 \frac{\aleph_{IS} }{T_{IS}} &= \frac{{{\aleph _{IS}^1} + \sum\limits_{k = 2}^{N^*} {{\aleph _{IS}^k}} }}{{\left\lceil {\frac{{f_3}}{{96}}} \right\rceil  \times {t_{tag}} + \left\lceil {\frac{{f_3}}{{96}}} \right\rceil  \times {t_w}}} \nonumber \\
  &\approx \frac{{96\left( {p_3{e^{ - p_3}} + p_3q{e^{ - p_3\left( {1 - q} \right)}} - p_3q{e^{ - p_3}}} \right)}}{{{t_{tag}} + {t_w}}}.
 \end{align}
 
To maximize the efficiency, we can calculate the derivative of the efficiency with respect to $p_3$. Here we set the length of $w$ to 96 to verify more tags in a slot. Owing to the existence of $q$, which is unknown by the reader in advance, we should first estimate the number of missing tags. Although we can use some existing methods to estimate the cardinality of tags before executing the ISMTI, the time overhead and protocol complexity will increase accordingly. Therefore, we propose a cardinality estimation method based on the replies of all the tags in each round of ISMTI.
Recall that the reader can obtain the expected  vector ${EV}$ by using its stored IDs and the actual vector ${AV}$ based on the replies from tags. Hence, we have: $\exists i \in \left[ {1...{f_3}} \right],{EV}[i] \ne {AV}[i]$. Let ${P_{11}}$ denote the probability that the current bit of a string is arranged to one present tag, we have
 \begin{align}
 {P_{11}} &= \left( {\begin{array}{*{20}{c}}
   {{N^*} - {M^*}}  \\
   1  \\
\end{array}} \right)\left( {\frac{1}{{f_3}}} \right){\left( {1 - \frac{1}{{f_3}}} \right)^{N - 1}} \nonumber\\
  &\approx \frac{{{N^*} - {M^*}}}{{f_3}}{e^{ - \frac{N}{{f_3}}}}
 \end{align}
where  ${M^*}$ is the number of unidentified missing tags in the current round.
Let ${N_{11}}$ denote the number of bits mapped by one present tag in each round, which follows the distribution of Bernoulli $\left( {{f_3},{P_{11}}} \right)$. Then we can calculate the expectation of the ${N_{11}}$ as 
 \begin{align}
{E_{{N_{11}}}} = {f_3} \times {P_{11}} = \left( {{N^ * } - {M^ * }} \right) \times {e^{ - \frac{{{N^ * }}}{{f_3}}}}.
  \end{align}

Let ${N_{1}}$ denote the number of bits arranged to one tag in each round. According to  (12), we  can calculate the expectation of the ${N_{1}}$ as
 \begin{align}
{E_{{N_{1}}}} = {f_3} \times {P_{IS}^1} =  {{N^ * }}  \times {e^{ - \frac{{{N^ * }}}{{f_3}}}}.
  \end{align}

Combing (27) and  (28), we can obtain the expressions of $M^*$ as 
\begin{align}
{M^* }= {N^ * } - \frac{{{E_{{N_{11}}}}}}{{{E_{{N_1}}}}} \times {N^ * }.
\end{align}

Substituting ${{N_1}}$ for ${E_{{N_1}}}$ and ${{N_{11}}}$ for ${E_{{N_{1}}}}$, we can calculate the estimators $\widehat{M^*}$ as 
\begin{align}
\widehat{{M^* }} = {N^ * } - \frac{{{N_{11}}}}{{{N_1}}} \times {N^ * }.
\end{align}

Therefore, we can obtain the missing rate $q$ of each round as 
\begin{align}
{q = \frac{{\widehat{{M^*}}}}{{{N^*}}}}.
\end{align}

Note that we cannot directly provide a closed-form expression of ${f_3}$ because $q$ is different in different scenarios. However, when $q$ is estimated, we can equate the derivative of ${\theta}$ to $0$ and calculate the optimal value of ${p_{opt}}$. Then, we can obtain the optimal value of ${f_3}$ as ${f_{opt}} = {{{N^ * }} \mathord{\left/
 {\vphantom {{{N^ * }} {q}}} \right.
 \kern-\nulldelimiterspace} {p_{opt}}}$.
\section{Discussion}
\subsection{Practical Implementation Issues}
To implement the proposed protocols on commercial RFID devices, we need to make the following modifications to the reader and tags. First, the tags need to execute hash calculations based on their ID and the random seed broadcast by the reader. Recently, such on-tag hashing has been implemented on commercial RFID devices using the fundamental and mandatory functionalities in the EPC C1G2 standard \cite{hashshixian}. Second, the tags need to add a counter to store the slot index. Compared with the overall circuit scale of modern tags \cite{5929555}, the added counter is negligible. Moreover, the functionality of recording the slot index is also used in previous protocols \cite{Liu2014Completely, PCMTI, CLS, Chen2017Efficiently}.
Third, the tags need to perform simple arithmetic operations. Compared with the EPC C1G2 protocol, the additional operation of our protocol is only the addition operations, and each passive tag is equipped with about 4000-5000 NAND gates. Therefore, it is feasible to deploy our protocol on passive tags \cite{IPP}.  On the other hand, to achieve the functionality of transmitting the indicator vector and the parameters in each round, the RFID reader needs to add extra custom commands supported by the EPC C1G2 protocol \cite{DFSA}. However, this operation does not need to change the original hardware circuit of the reader. Therefore, the proposed protocols have the potential to be implemented on commercial RFID systems.

\subsection{Multiple-Reader RFID Systems}
Due to the limited interrogation range, which is normally less than 10 meters \cite{Liuxiu}, it is challenging for a single RFID reader to cover a large-scale identification area. Although handheld readers are available to scan a specific area of a large system periodically, multiple readers are also required in many scenarios. However,
if two or more adjacent readers simultaneously interrogate the tags in a multiple-reader RFID system, the readers' signals will collide. In this case, the overlapping tags cannot receive correct commands.   
To avoid the collision of readers, some advanced reader-scheduling algorithms \cite{duxieqi1, duxieqi2} were proposed to determine the query order of each reader. 
Since reader scheduling strategy is not the key point of this paper, we consider using the Colorwave scheme \cite{color} to illustrate the extension of our protocols in multiple-reader RFID systems. Specifically, we use one color to mark readers without collision as many as possible. Then, we mark the readers without collision among the remaining uncolored readers using a different color. This coloring process is executed continuously until all readers are marked. Finally, we active all readers of the same color to execute the proposed protocols in parallel.

In multiple-reader RFID systems, the back-end server stores the IDs of all the tags, but each reader does not have the tag IDs in its interrogation range. Therefore, we use the whole tag set as input to calculate the same parameters and indicator vectors for the first batch of scheduled readers.
When receiving the replies from tags, each reader sends the information to the back-end server for data aggregation. After all the readers having completed the missing tag identification, we activate readers of another color to restart the protocol. To save execution time, the back-end server deletes the identified present tags from the inventory list and then uses IDs of the remaining tags to calculate the parameters and vectors for the next batch of scheduled readers.
The scheduled process is continuously executed until all readers have completed the missing tag identification.

\begin{figure*}[htbp]
    \centering
    \subfigure[Execution time of SSMTI when varying the length $w$ from 1 to 96]{
        \includegraphics[width=0.275\textwidth]{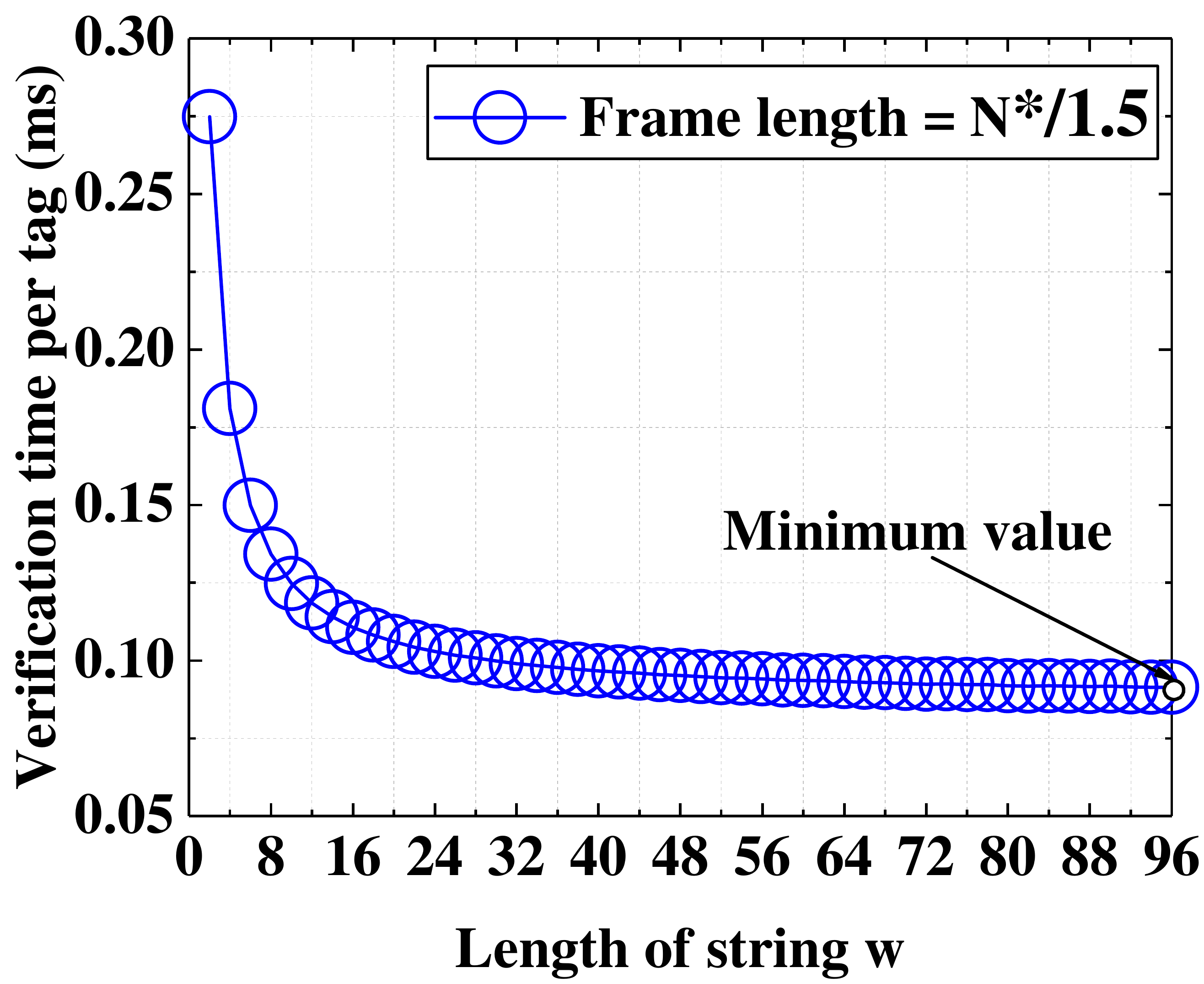}
    }
    \subfigure[Execution time of SSMTI when varying the $p$ from 0.5 to 4]{
        \includegraphics[width=0.275\textwidth]{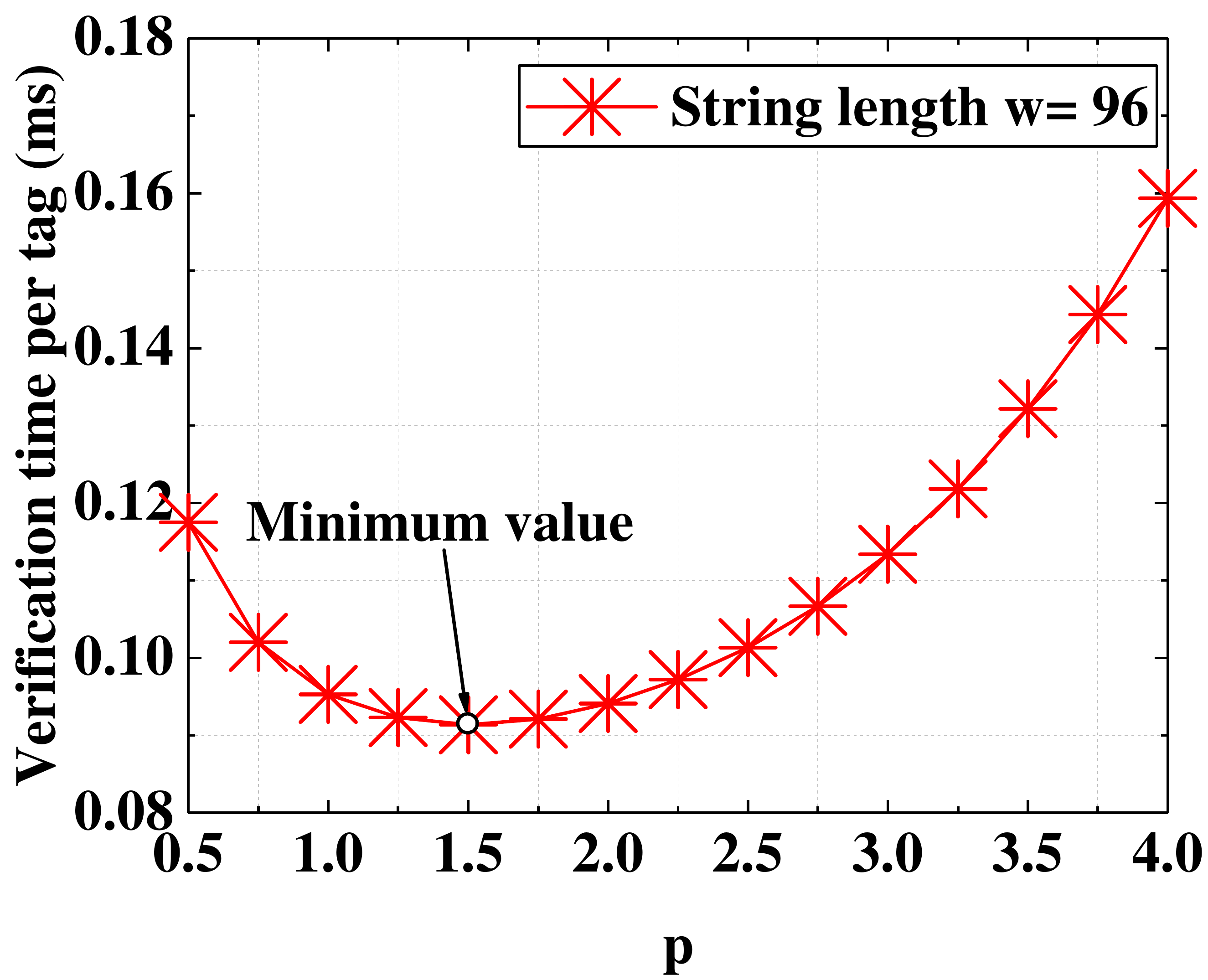}
    }
    \subfigure[Execution time of ISFMTI when varying the $p$ from 0 to 26]{
        \includegraphics[width=0.27\textwidth]{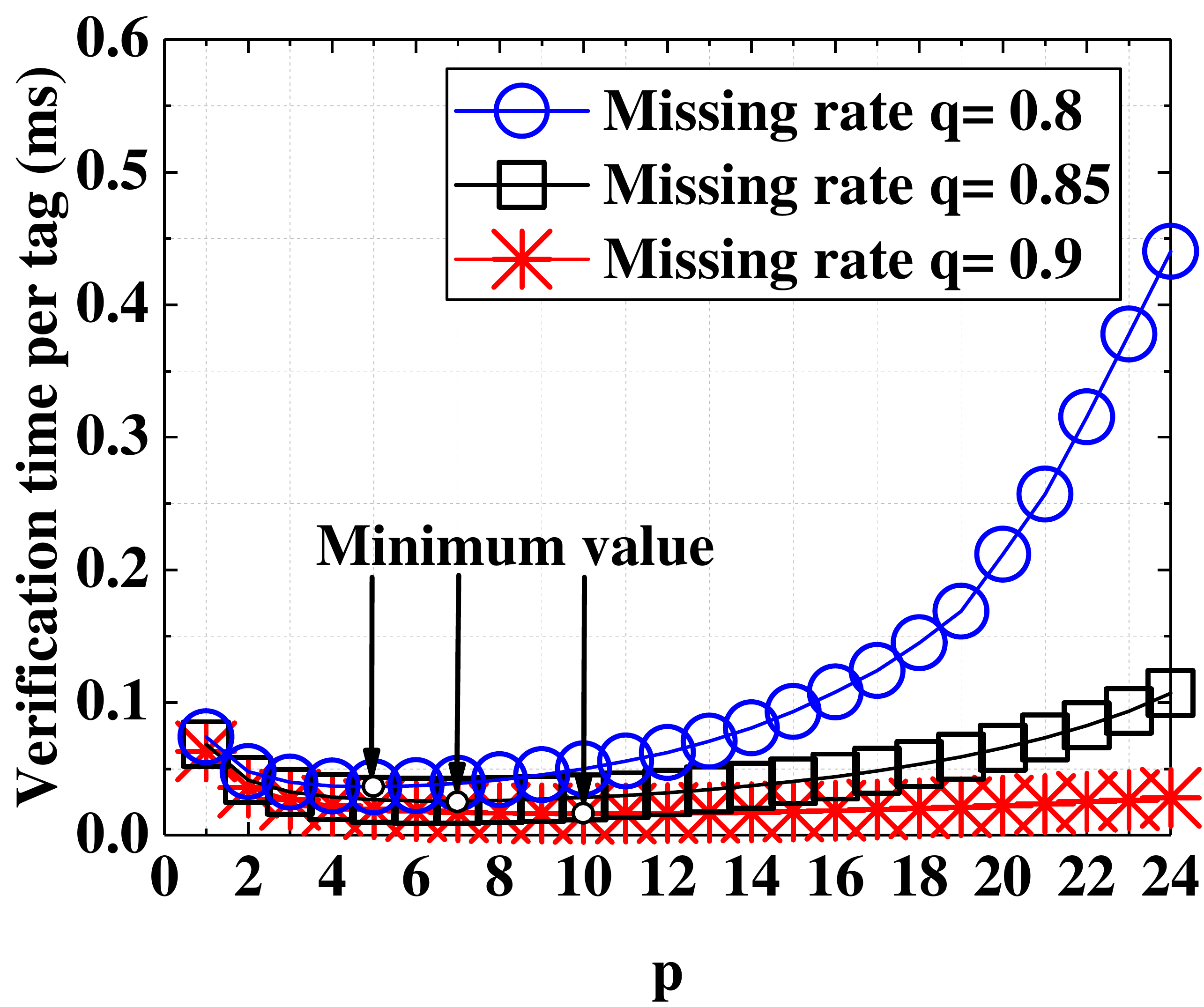}
    }
    \caption{Impact of string length and frame length on execution time}
    \label{fig7}
\end{figure*}

\begin{figure*}[htbp]
    \centering
    \subfigure[Execution time of different approaches when varying the number of candidate tags from $1,000$ to $10,000$]{
        \includegraphics[width=0.28\textwidth]{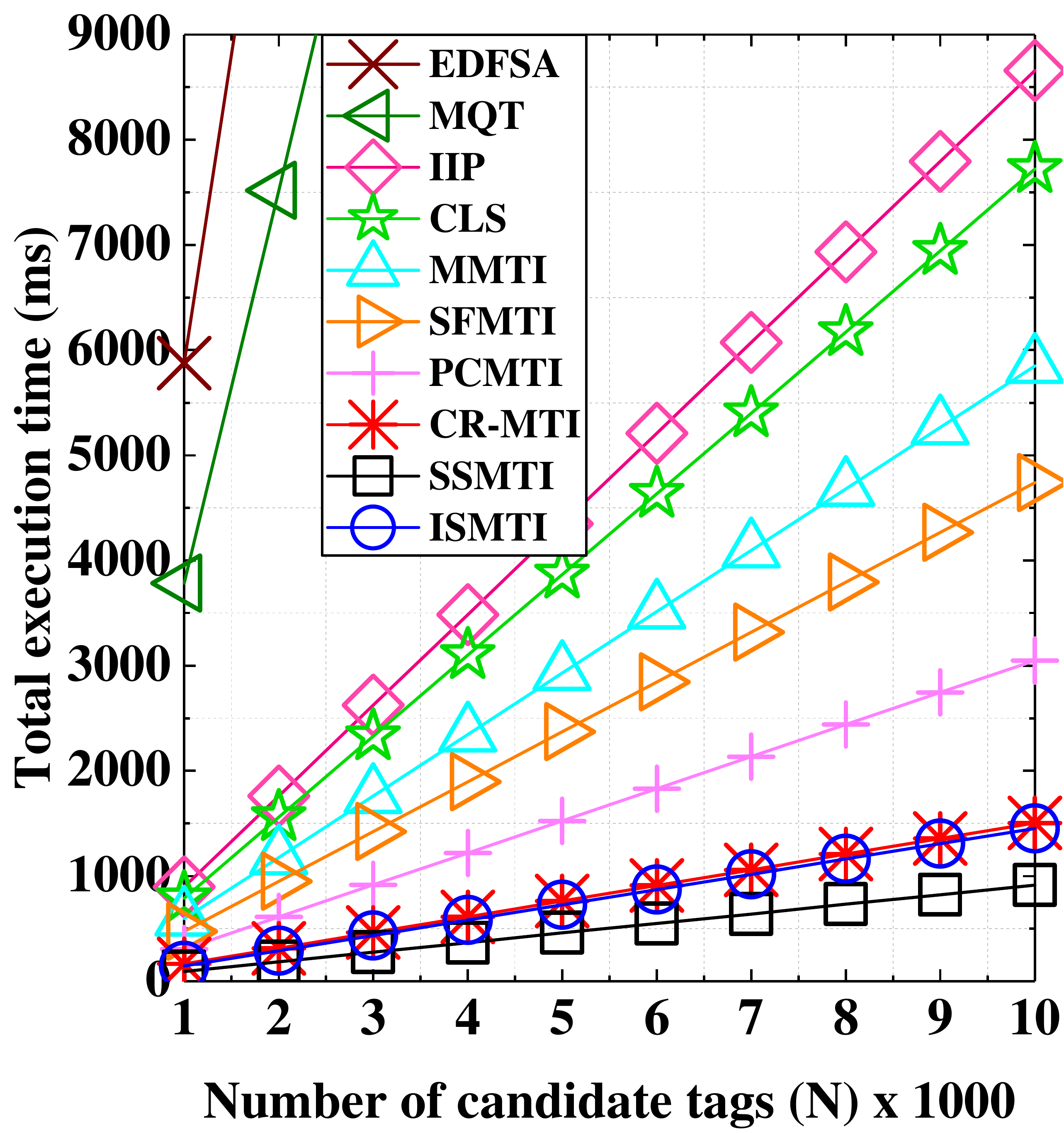}
    }
    \subfigure[Number of message bits transmitted by reader and tags when varying the number of candidate tags from $1,000$ to $10,000$]{
        \includegraphics[width=0.28\textwidth]{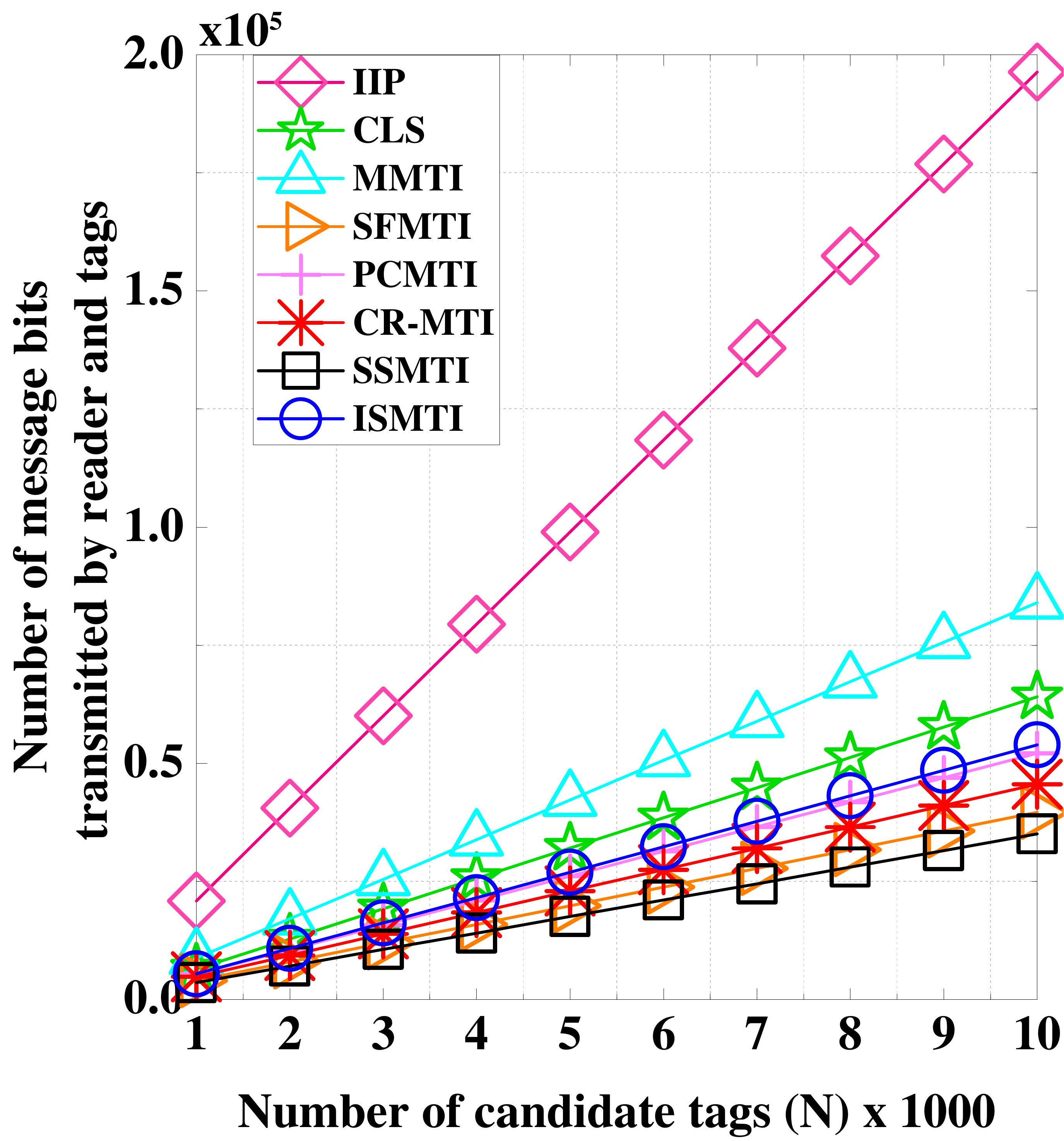}
    }
    \subfigure[Number of message slots for tag message transmission when varying the number of candidate tags from $1,000$ to $10,000$]{
        \includegraphics[width=0.286\textwidth]{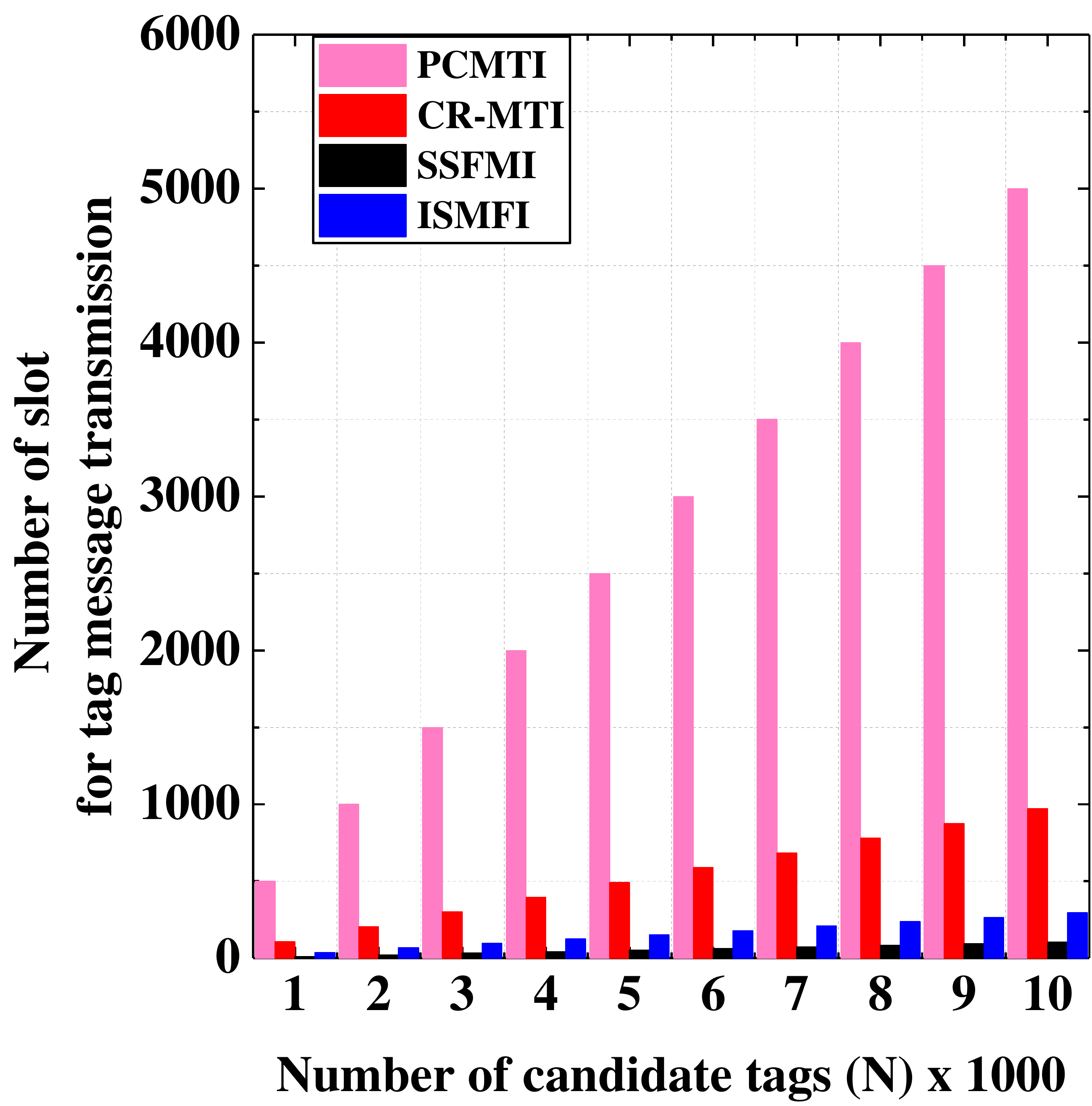}
    }
    \caption{Performance comparison when varying the number of candidate tags}
    \label{fig8}
\end{figure*}
\section{Performance Evaluation}
In this section, we evaluate the performance of the proposed SSMTI and ISMTI protocols. For a fair comparison, we use the same
wireless communication settings for each protocol based on the specification of the Philips I-Code system \cite{feilipu}. Specifically, after including the required waiting times between two transmissions, the time for the reader or tags to transmit a 1-bit message is ${t_s} = 0.4$ ms, and the time to transmit a 96-bit tag ID is ${t_{tag}} = 2.4$ ms. In other words, the data rate is about $96/2.4=40$ kbps \cite{MMTI, PCMTI}. Therefore, it takes ${0.4+(w-1)\times 0.025}$ ms to transmit a $w$-bit string for each tag. In addition, the communication channels between the reader and tags are considered as error-free as in the literature \cite{IIP, MMTI, PCMTI, CLS}. For reliability, we carry out five hundred experiments  for each parameter group and report the averaged results.

\subsection{Comparative Works}
\textcolor {black} {To verify the efficiency of the proposed protocols, we consider two types of tag identification protocols as benchmarks. The first type is anti-collision protocol, which consists of a well-known Aloha-based protocol, namely Enhanced Dynamic Framed Slotted Aloha (EDFSA) \cite{EDFSA}, and an advanced tree-based scheme, which is M-ary Query Tree scheme (MQT) \cite{shu}. On the other hand, the second type is advanced missing tag identification protocols, including IIP \cite{IIP}, MMTI \cite{MMTI}, SFMTI \cite{Liu2014Completely}, CLS \cite{CLS}, PCMTI \cite{PCMTI}, and CR-MTI \cite{Suchao}.} 
 
 We set the frame length of EDFSA, IIP, MMTI, SFMTI, PCMTI to ${N^ * }$ \cite{EDFSA}, ${N^ * }/1.1516$ \cite{IIP}, ${N^ * }$\cite{MMTI}, ${N^ * }/1.68$ \cite{Liu2014Completely},  $0.8735{N^ * }$ \cite{PCMTI}, and
${N^ * }/34$ \cite{Suchao}, 
 where ${N^*}$ is the number of tags not verified by the reader at the beginning of each round. For CLS, its frame length is set according to the missing rate of each round \cite{CLS}.
\subsection{Impact of Frame Length and String Length}
In this section, we simulate the proposed SSMTI and ISMTI under different parameters and compare the average execution time to identify a candidate tag. 
Here we set the number of candidate tags $N$ as $10,000$.
Based on the analysis in Section IV, the optimal frame length of SSMTI is  $ {f_{SS}} = {{{N^ * }} \mathord{\left/
 {\vphantom {{{N^ * }} {1.5}}} \right.
 \kern-\nulldelimiterspace} {1.5}}$. Besides, the string length of SSMTI is  $ {w_ {SS}} = 96$. On the other hand, the string length of ISMTI is $ {w_ {IS}} = 96 $, and the optimal frame length ${f_{IS}}$ is set based on the missing rate.  In Fig. 5(a), we vary the parameter $w$ of  SSMTI. Similarly, to verify the setting of $f$, we also vary the parameter $p=\frac{{{N^ * }}}{f}$ of SSMTI and ISMTI in Fig. 5. We can observe that the identification time of THSMTI and SSMTI are minimized when ${w_{SS}}=38$ and ${f_{SS}}={N^*}/1.5$.  
We can also observe in Fig. 5(c) that there is an optimal value of $p$ in ISMTI under different missing rates.  Therefore, the simulation results match our theoretical analysis.  
\subsection{Impact of Number of Candidate Tags}
\textit{Execution time}:  According to existing works
\cite{Liu2015Unknown, Liu2015Sampling, Liu2014Completely, MMTI, PCMTI, CLS}, the execution time is one of the most critical performance metrics in missing tag identification. The execution time 
can be obtained by calculating the time required by the reader to transmit the indicator vectors and the time required by tags to reply. Note that the construction of indicator vectors and most of the calculation tasks are done on the reader side, which has powerful calculation and storage capabilities. Moreover, these tasks can be performed by the reader offline. Therefore, the execution time on the reader side is ignored like other protocols\cite{IIP, MMTI, Liu2014Completely, PCMTI, CLS}.

        In Fig. 6(a), we compare the execution time of the proposed protocols with existing works. Here we set the missing rate $q$ as $0.1$ and vary the number of candidate tags $N$ from $1,000$ to $10,000$. We can observe that the execution time of all the protocols increases as the number of candidate tags grows. The reason is that more time slots are required by the reader to verify more candidate tags. The performance of the proposed SSMTI and ISMTI are better than other protocols. For example, when the number of candidate tags $N=10,000$, the total execution time of EDFSA, MQT, IIP, SFMTI, PCMTI, CLS, and \textcolor{black} {CR-MTI is approximately $58.75$ s, $37.61$ s,  $8.66$ s, $4.73$ s, $3.05$ s, $7.72$ s, and $1.51$ s.} \textcolor{black} {However, the execution time of  ISMTI and SSMTI is $1.45$ s, and $0.91$ s. \textcolor{black} {Compared with the state-of-the-art CR-MTI protocol, the proposed SSMTI protocol reduce the execution time by $39.74\%$.} The underlying reasons are as follows.  First,  although EDFSA and MQT can reduce tag collisions, these protocols require each tag to transmit ID multiple times. Therefore, it is time-consuming to use anti-collision protocols to solve the problem of missing tag identification. On the other hand, most tag identification protocols, such as IIP, MMTI, and SFMTI, can only identify one tag in each time slot. The proposed protocols use bit tracking technology to verify multiple tags together in each collision slot, thus accelerating the identification process. Moreover, different from PCMTI and CR-MTI, which can only use part collision slots for tag verification and suffer from serious bit waste in each string, SSMTI exploits all collision slots and makes each bit of string correspond to an unverified tag, thus fundamentally improving the identification efficiency.}  
  
 \begin{figure*}[htbp]
    \centering
    \subfigure[Execution time of different approaches when varying the missing rate of tags from 0.1 to 0.95]{
        \includegraphics[width=0.34 \textwidth]{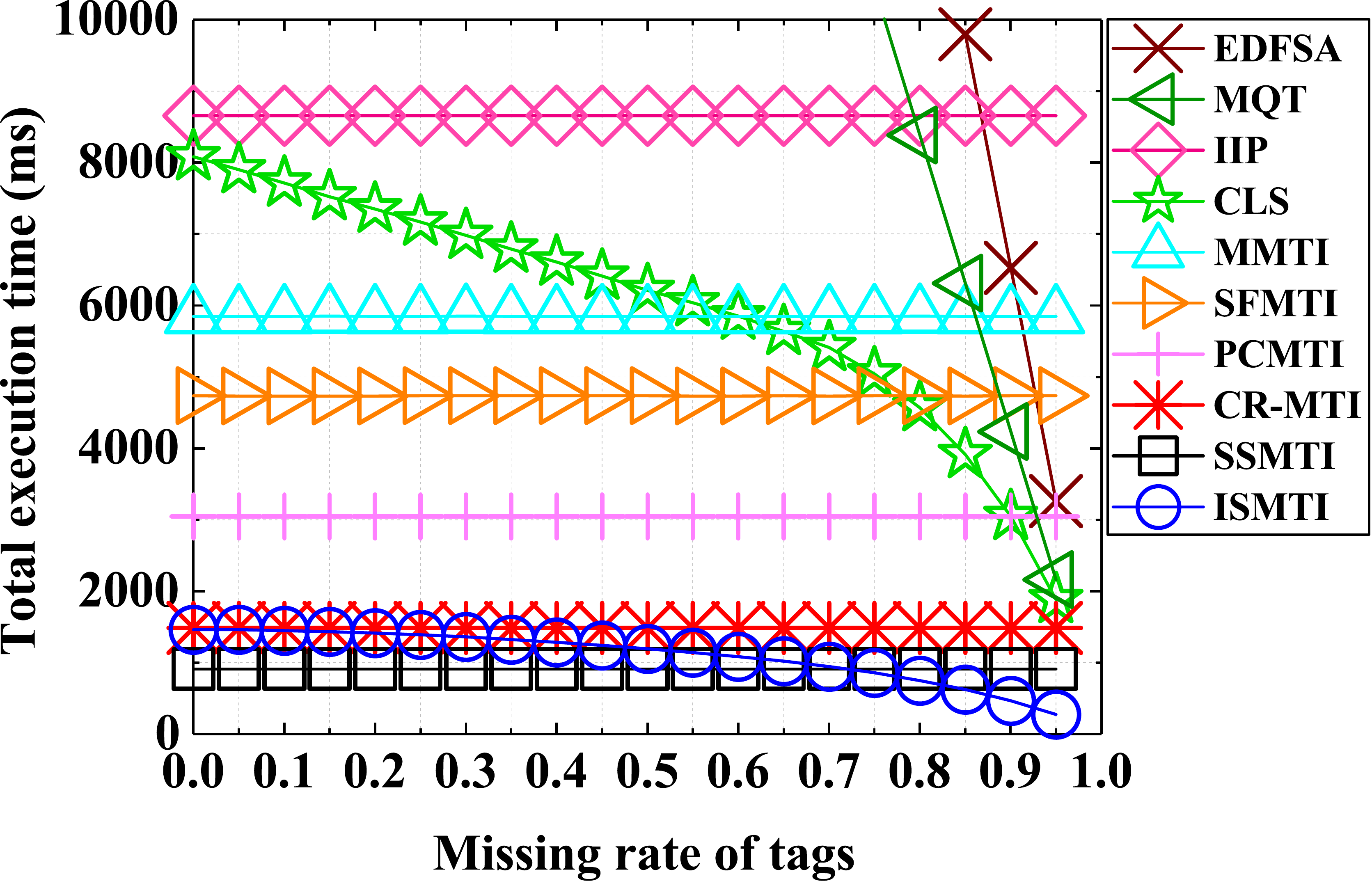}
    }
    \subfigure[Execution time of different approaches when varying the missing rate of tags from 0.994 to 0.999]{
        \includegraphics[width=0.28\textwidth]{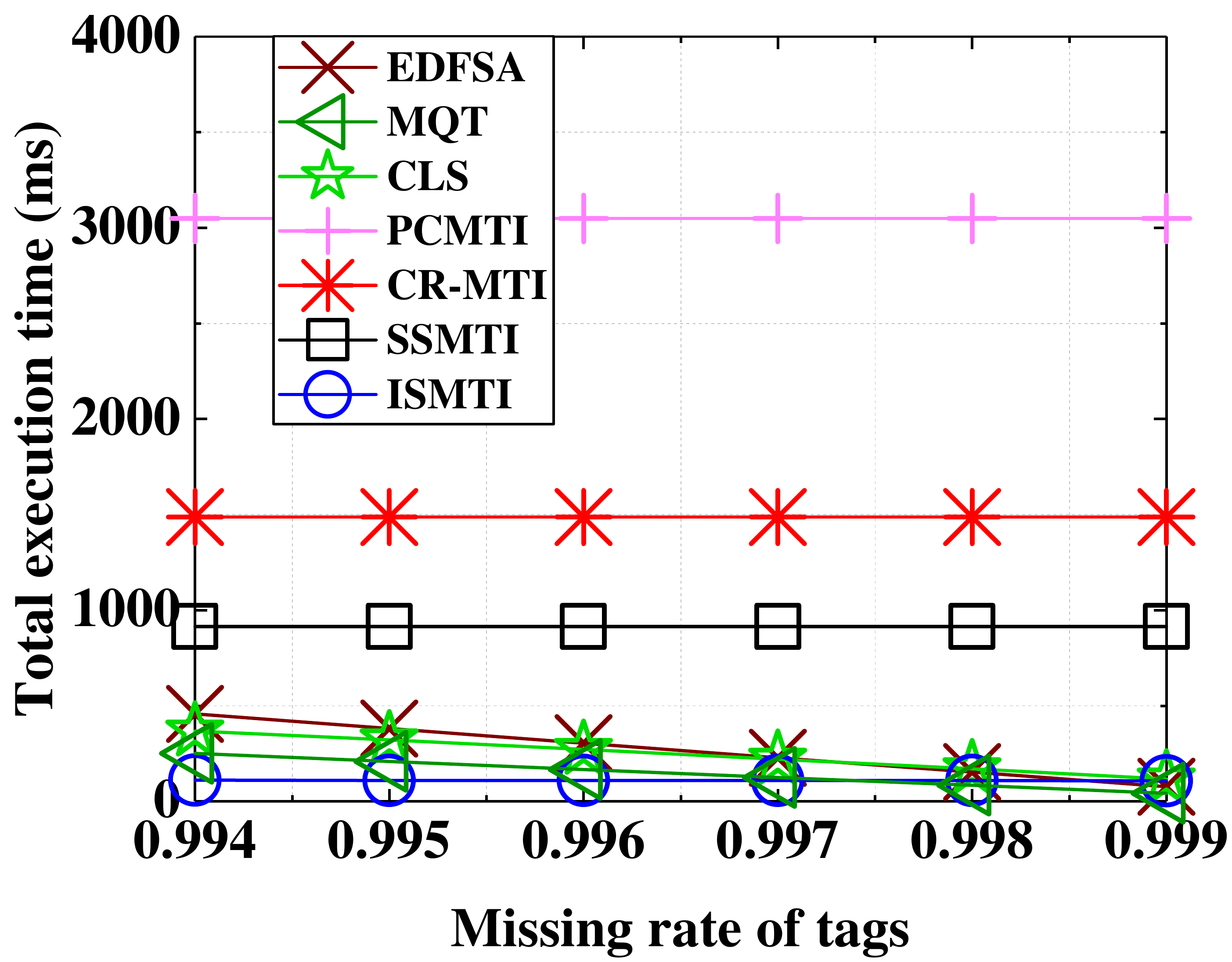}
    }
    \caption{Performance comparison when varying the missing rate of candidate tags}
    \label{fig9}
\end{figure*}
 \textit{Number of bits transmitted}: The number of bits transmitted by the reader and tags can reflect the complexity of the communication and implies the energy consumption of the communication process \cite{Subit}. Under the same settings of Fig. 6(a), we compare the number of bits transmitted by the reader and tags in Fig. 6(b). We can observe that CR-MTI has to transmit more bits than SSMTI when the missing rate is low. The reason is that CR-MTI uses the hash method to arrange the bit positions in the string for each tag, resulting in some bits of the string may not be arranged to any tags. On the other hand, SSMTI transmits the fewest bits. The underlying reasons are as follows. First, the coding method of the indicator vectors in SSMTI reduces the bit transmitted by the reader. Second, SSMTI arranges each tag a unique bit in the string to reply, thus avoiding the waste of bits in CR-MTI.

 \textit{Number of slots transmitted}: Fig. 6(c) shows the number of actual response slots required by the reader to verify all tags. Compared with IIP, MMTI, and SFMTI, PCMTI can verify two tags simultaneously in a slot, thus requiring a smaller number of response slots. Based on the bit tracking technology, the proposed protocols can verify more than two tags in a slot. Therefore, SSMTI, and ISMTI require fewer response slots than PCMTI. Moreover, among the proposed protocols, SSMTI and ISMTI map more tags in each slot than CR-MTI, so they require fewer response slots than the CR-MTI protocol.
\begin{figure}
\centerline{\includegraphics[width=0.6\columnwidth]{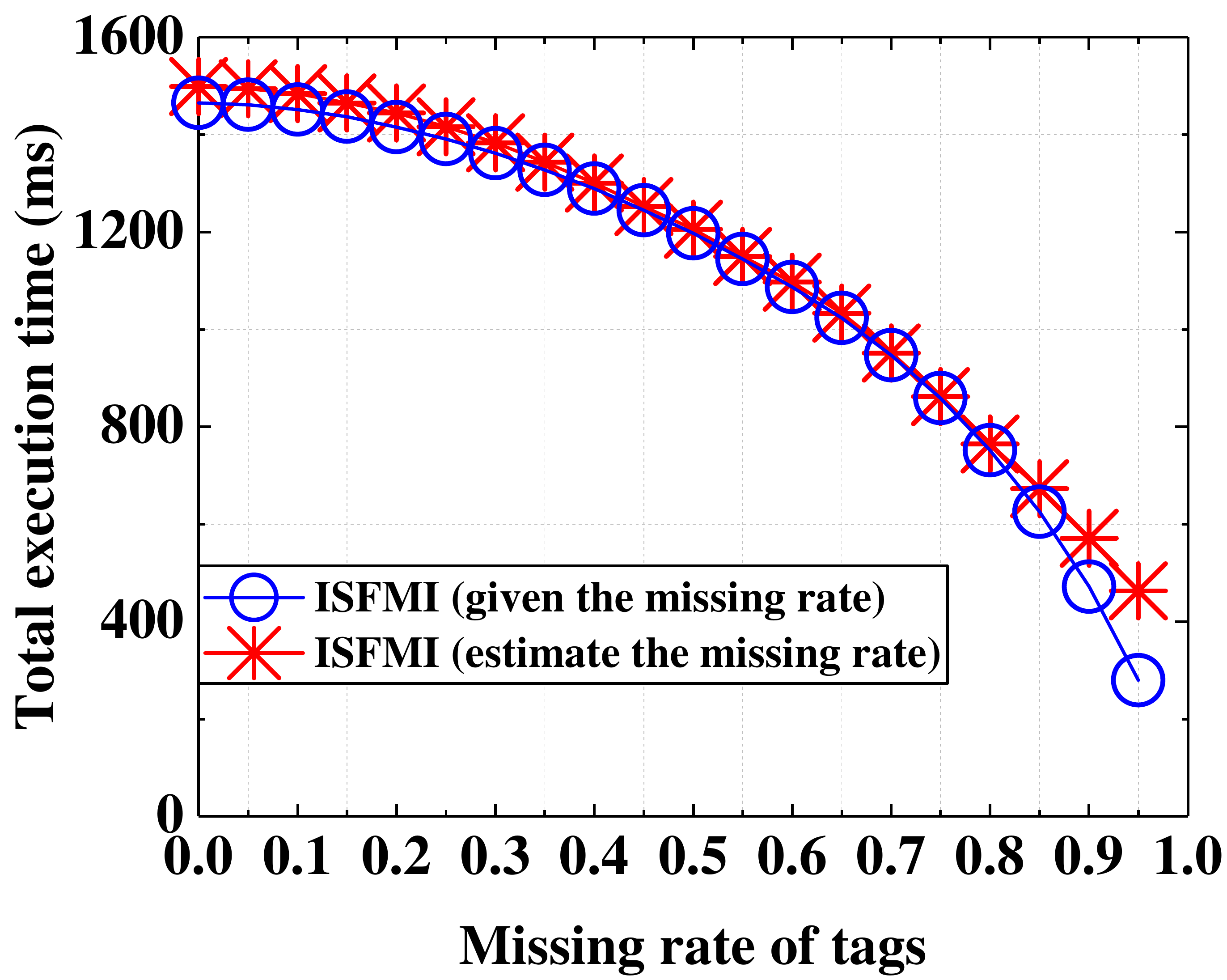}}
\caption{Impact of estimation error on the execution time of ISMTI }
\label{fig10}
\end{figure}
\subsection{Impact of Missing Rate}
In Fig. 7, we set $N=1,0000$, and vary the missing rate to evaluate the proposed protocols. 
We can observe in Fig. 7(a) that the execution time of IIP, MMTI, SSFMTI, and PCMTI is stable. The reason for this result is that these protocols only verify one or two tags in a slot. Moreover, the execution time of CR-MTI, SSMTI is also stable because each tag is arranged to one bit in the string. Therefore, the execution time of the above protocols is only affected by the number of candidate tags. On the other hand, the execution time of CLS gradually decreases as the missing rate increases. The reason for this result is that CLS checks the expected collision slots for tag verification. When the missing rate is high, most expected collision slots will turn out to be empty during actual execution. However, when the missing rate is low, the performance of CLS is worse than SFMTI because CLS cannot verify the collided tags in a slot. The proposed protocols run faster than existing approaches when the missing rate is less than $0.95$. Moreover, as the missing rate increases, the execution time of the ISMTI protocol decreases quickly, so that ISMTI gradually becomes the fastest. For example, when the missing rate is $0.9$, the execution time of CLS is approximately $3.05$ s. \textcolor {black} {ISMTI consumes $0.47$ s, which reduce the execution time by $84.59\%$, $68.87\%$, and $48.36\%$ compared with CLS, CR-MTI, and SSMTI.} 
Unlike CR-MTI and SSMTI that do not consider the impact of missing rate on execution time, ISMTI arranges multiple tags to the same bit in the string at high missing rates. In this way, the reader in ISMTI can verify more tags in each string than SSMTI, thereby further reducing the execution time.  \textcolor {black} {Besides, ISMTI can adaptively adjust the string design strategy according to the changing missing rate, thus avoiding the performance degradation of CLS.}

In Fig. 7(b), we vary the missing rate from $0.994$ to $0.999$. We can observe that the execution time of EDFSA and MQT is also reduced because the number of present tags in the reader's interrogation region is small. Moreover, the performance of EDFSA and MQT is better than that of CLS when most tags are missing. However, EDFSA and MQT are time-consuming in most scenarios and not suitable for identifying some valuable items because tag IDs are directly transmitted by tags in the air \cite{Chen2017Efficiently}.

\subsection{Impact of Estimation Error}
\begin{figure}[htbp]
    \centering
\subfigure[The percentage of misidentified tags when the missing rate is 0.1]{
        \includegraphics[width=0.29\textwidth]{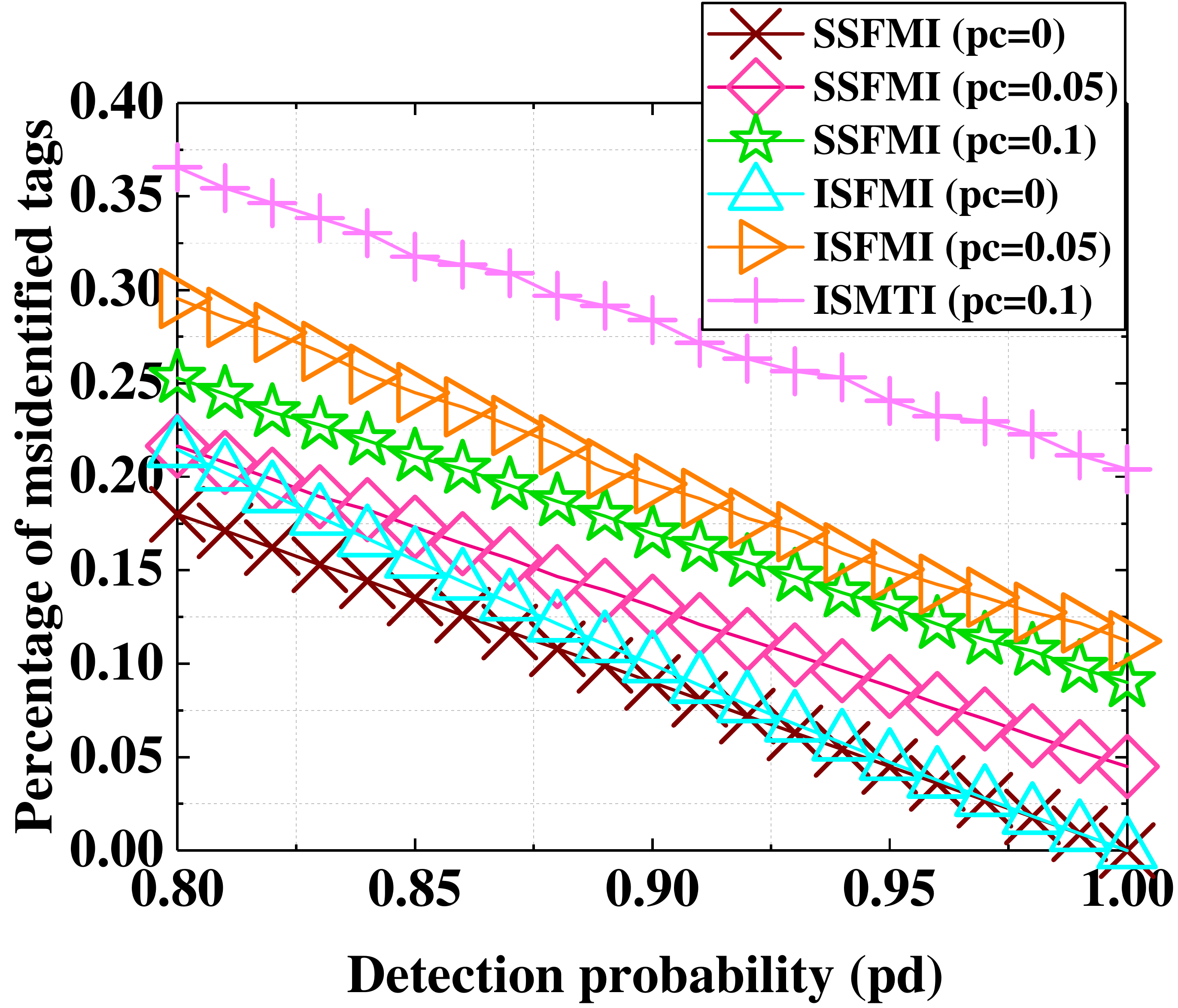}
    }
    \subfigure[The percentage of misidentified tags when the missing rate is 0.1]{
        \includegraphics[width=0.29\textwidth]{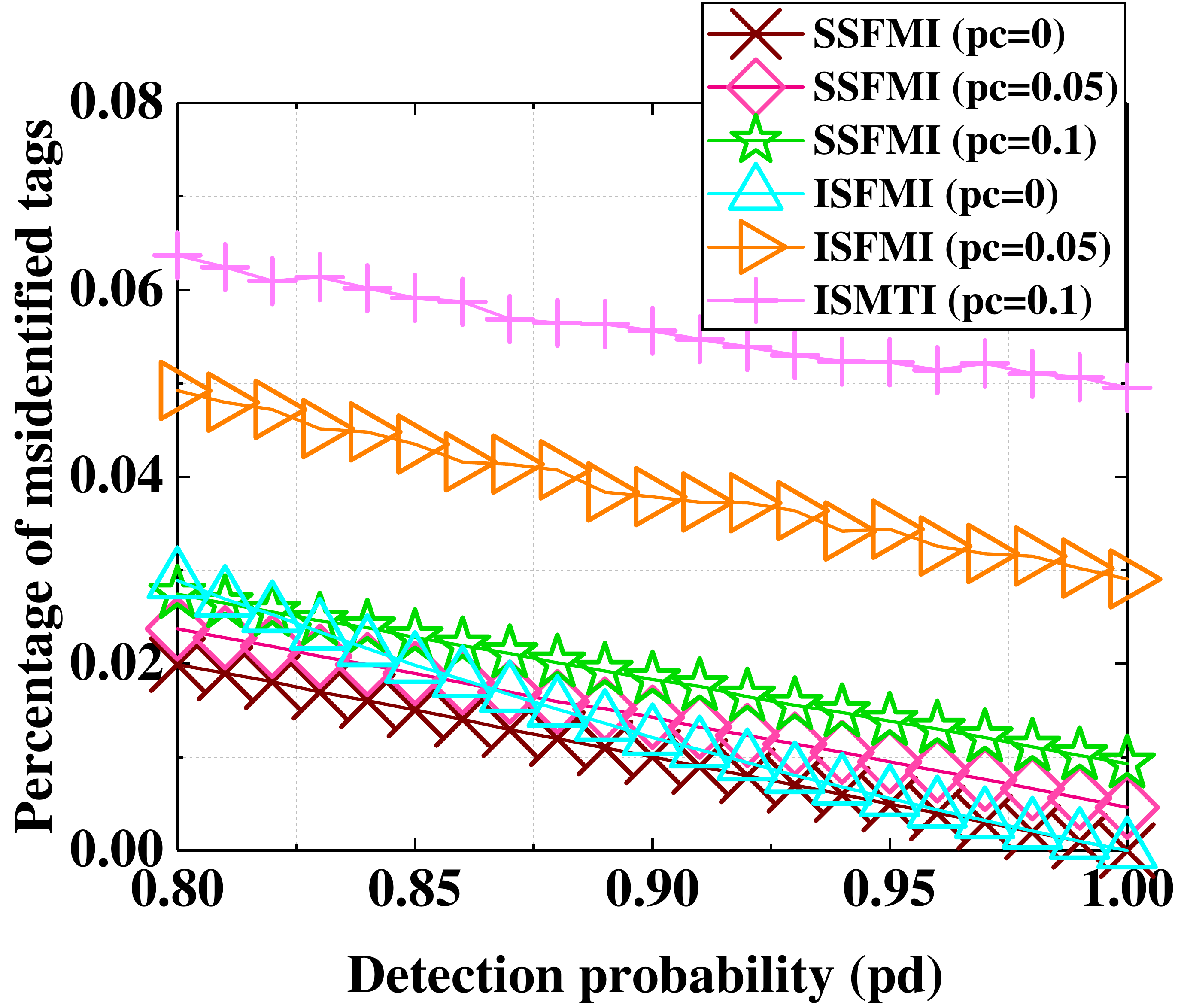}
    }
    \caption{Evaluating the identification accuracy of the proposed protocols under
imperfect channel conditions}
    \label{fig11}
\end{figure}
When the reader does not know the missing rate, the ISMTI protocol can use the method in Section VI to estimate the number of missing tags. In Fig. 8, we compare the execution time of ISMTI using the proposed estimation method with ISMTI that is given the missing rate.  Here we set the initial estimation of the missing rate $p$ to $0.5$ in the first round of ISMTI (estimate the missing rate). We can observe in Fig. 8 that the execution time of ISMTI (estimate the missing rate) is close to ISMTI (given the missing rate) when the missing rate ranges from $0$ to $0.9$. Therefore, when the missing rate is not given, ISMTI can achieve the expected performance. Besides, ISMTI does not need to add additional estimation protocols.

\subsection{Impact of Channel Error}
In the above analysis, the communication channel is considered to be error-free as in existing works \cite{Liu2014Completely, CLS, Suchao}. However, channel errors may happen in many realistic scenarios. 
Similar to existing works that adopt bit-tracking technology \cite {zhang1, zhang2}, the performance of SSMTI and ISMTI are mainly influenced by detection errors and capture effect. Specifically, detection errors are caused by the weak backscattering signal from some passive tags. In this case, the reader cannot tag signal and therefore misidentify these tags as missing. On the other hand, the difference in the distance between tags and the reader causes the signal strength of each tag to be different. When the signal strength of a tag is significantly stronger than other tags in a time slot, the reader can only detect this tag, thus causing the capture effect. 
Note that we do not consider bit errors like other RFID identification protocols \cite{zhang1, zhang2}. The reason is that in most indoor scenes where the distance between the reader and the tag is often less than $10$ m, the probability of bit errors is less than $10^-6$ \cite{zengyi}, which has a negligible impact on tag identification \cite{zhang1, zhang2}. 

\textcolor{black} {In Fig. 9, we set the missing rate to $0.1$ in Fig. 9(a) and $0.9$ in Fig. 9(b) to evaluate the proposed protocols. Here the number of
candidate tags $N$ is $10,000$. We can observe that the identification accuracy of the proposed protocols decreases when the capture effect and detection errors occur. This is because multiple existing tags in SSMTI and ISMTI transmit strings in the same time slot. When the capture effect happens, the reader can only identify one present tag and treat other tags as missing. Besides, the detection errors can also cause some present tags to be not detected by the reader. To improve identification accuracy, the reader can reduce the string length in each slot. Moreover, it is worth noting that the reader usually executes the identification protocol periodically. Therefore, these undiscovered tags have a high probability of being detected by the reader in the subsequent cycle. We can also observe that the identification accuracy of SSMTI and ISMTI becomes higher when the missing rate is high. This is because most actual time slots are empty slots, which are not affected by capture effects and detection errors. By using the proposed protocol in this scenario, the reader can identify multiple missing tags at once to significantly improve the identification efficiency and ensure high identification accuracy.}

\section{Conclusion}
We investigated a critical problem of missing tag identification under different missing rates. Unlike most protocols that verify tags one by one in singleton slots, we exploit collision slots for tag verification and proposed three efficient missing tag identification protocols based on bit tracking technology. We first proposed the Sequential String based Missing Tag Identification (SSMTI) protocol, which converts all time slots to collision slots and enables multiple tags in each slot to simultaneously reply to a designed string. By adopting the bit tracking technology to decode the combined string, the reader can verify multiple tags together. 
To further accelerate the identification process under high missing rates, we proposed the Interactive String based Missing Tag Identification (ISMTI) protocol. Unlike SSMTI that arranges each tag to a unique bit of the string, 
multiple tags in ISMTI can be arranged to the same bit in the string so that the reader can identify more tags using a shorter string than SSMTI. Besides, the verification mechanism in ISMTI can be dynamically adjusted according to the missing rates.
We also analyzed the parameter configuration of the proposed protocols to minimize the execution time. Numerical results show that the proposed protocols outperform state-of-the-art solutions in terms of communication complexity and time efficiency. In future work, we intend to investigate the rapid identification of some important tags using commercial RFID systems.

\ifCLASSOPTIONcaptionsoff
  \newpage
\fi



%
\bibliographystyle{ieeetr}
\bibliography{bare_jrnl}

%








\end{document}